\documentclass[submission, Phys]{SciPost}
\hypersetup{colorlinks,linkcolor={red!50!black},citecolor={blue!50!black},urlcolor={blue!80!black}}
\usepackage{amsmath,amssymb}
\usepackage[caption=false]{subfig}
\usepackage{tikz}
\usepackage{tikz-feynman}
\usepackage{slashed}
\usetikzlibrary{shapes.misc,arrows.meta,decorations.pathmorphing,shapes,arrows,positioning,graphs}

\begin{document}

\begin{center}{\Large\textbf{Universal transport at the composite Fermi liquid to Fermi liquid transition in moir\'e Chern bands\\}}\end{center}

\begin{center}
Youxuan Wang\textsuperscript{1}, Rongning Liu\textsuperscript{1}, Feng Liu\textsuperscript{1} and Xue-Yang Song\textsuperscript{1$\star$}\\[2pt]
{\bf 1} Department of Physics, The Hong Kong University of Science and Technology, Clear Water Bay, Hong Kong, China\\[2pt]
$^\star$ {\small\sf songxy@ust.hk}
\end{center}

\begin{center}\today\end{center}

\section*{Abstract}
{\bf
We study universal electrical transport at continuous transitions between a composite Fermi liquid and a Fermi liquid in moir\'e Chern bands at fillings $\nu=-1/2$ and $-3/4$. The critical theory couples Dirac fermions at a boson Laughlin--superfluid transition to a spinon Fermi surface through Chern--Simons gauge fields. We derive the Ioffe--Larkin rule for the full resistivity tensor and solve a quantum Boltzmann equation at leading order in a large-$N$ expansion. Spinon Landau damping adds a quasistatic scattering channel with weight independent of the spinon Fermi energy. At $N=1$ we obtain clean-limit resistivity jumps $\Delta\rho_{xx}\approx5.896\,h/e^2$ and $0.1274\,h/e^2$ at the two fillings.
}

\vspace{10pt}
\noindent\rule{\textwidth}{1pt}
\tableofcontents\thispagestyle{fancy}
\noindent\rule{\textwidth}{1pt}
\vspace{10pt}

\section{Introduction}
\label{sec:introduction}

moir\'e superlattices provide a highly tunable platform for strongly correlated electrons, as first demonstrated by the correlated insulators and superconductivity of twisted bilayer graphene~\cite{cao2018correlated,cao2018unconventional}.
In graphene- and transition-metal dichalcogenide (TMD)-based moir\'e systems, these correlations coexist with nontrivial band topology, and narrow, well-isolated Chern bands can be realized at zero magnetic field~\cite{sharpe2019emergent,serlin2020intrinsic,liQuantumAnomalousHall2021,Foutty_2024}.
Correlated and fractional Chern insulators~\cite{nuckolls2020strongly,xie2021fractional}, as well as integer and fractional quantum anomalous Hall states at zero field~\cite{serlin2020intrinsic,liQuantumAnomalousHall2021,caiSignaturesFractionalQuantum2023,zengThermodynamicEvidenceFractional2023,park2023observation,lu2023fractional}, have been observed in these systems and have stimulated extensive theoretical work on fractionalized phases in Chern bands~\cite{zhang2019nearly,ledwith2020fractional,repellin2020chern,abouelkomsan2020particle,wilhelm2021interplay,wu2019topological,yu2020giant,devakul2021magic,li2021spontaneous,crepel2023fci,song_2024_intertwined,zhang2024moore,song_2024density,zhou2024fractional,dong2024anomalous,dong2024theory,shi2025doping,kim2025topological,luiz1,luiz2}.
Beyond gapped topological orders, numerical studies of valley-polarized twisted MoTe$_2$ find a composite Fermi liquid (CFL) at lattice fillings $\nu=-1/2$ and $\nu=-3/4$~\cite{CompositeFermiLiquiddong2023,ZeroFieldCompositeFermigoldman2023}, and the magnitude of the Hall resistivity measured near $\nu=-1/2$ is close to $2h/e^2$, the value of the half-filled lowest Landau level~\cite{park2023observation}.
These results support the view that an isolated Chern band with unit Chern number can closely mimic the lowest Landau level.

moir\'e materials differ from conventional Landau-level systems in their in situ tunability.
A perpendicular displacement field changes the effective bandwidth while keeping the interaction scale comparatively fixed~\cite{liQuantumAnomalousHall2021,park2023observation}, electric-field-tuned transitions out of Chern insulators have been observed~\cite{caiSignaturesFractionalQuantum2023,zengThermodynamicEvidenceFractional2023}, and bandwidth-tuned Mott transitions have recently been realized~\cite{xia2026}.
Bandwidth-driven quantum phase transitions at fixed filling are therefore experimentally accessible.
The transition between the CFL and a Fermi liquid (FL) is a particularly interesting example, since the competition between these two phases is expected to be tunable by the displacement field~\cite{CompositeFermiLiquiddong2023}.
Such a transition raises a fundamental question of how a compressible non-Fermi liquid of composite fermions, with a nontrivial topological response, can evolve continuously into an ordinary metal of electron-like quasiparticles.
Continuous CFL--FL transitions were proposed within a projective construction of the CFL~\cite{ContinuousTransitionsCompositebarkeshli2012}.
Building on this construction, Ref.~\cite{song_2024phase} developed parton field theories for bandwidth-tuned transitions in moir\'e Chern bands and showed that, owing to filling constraints, translation symmetry alone suffices to enable a continuous CFL--FL transition.
In this description the electron is a product of a charged boson and a neutral fermion bound by an emergent $U(1)$ gauge field~\cite{WeakMagnetismNonFermisenthil2004,TheoryContinuousMottsenthil2008a}.
The neutral fermions form a spinon Fermi surface on both sides of the transition, whereas the boson forms a Laughlin state in the CFL and condenses in the FL.
The CFL--FL transition is thus a Laughlin--superfluid transition of the charged boson~\cite{PhysRevB.89.235116} in the presence of a neutral Fermi surface.
Its critical theory, constructed for both $\nu=-1/2$ and $\nu=-3/4$~\cite{song_2024phase}, consists of Dirac fermions coupled to emergent gauge fields with Chern--Simons terms~\cite{MottTransitionAnyonchen1993,TransitionsQuantumHallwen1993,EmergentMultiflavorQED3lee2018}, and these gauge fields are in turn coupled to the spinon Fermi surface.

Electrical transport is a direct and readily accessible probe of this criticality.
Reference~\cite{song_2024phase} identified its most striking signature as a universal jump of both the longitudinal and Hall resistivities, $\rho_{xx}$ and $\rho_{xy}$, at the critical point, analogous to the universal resistivity jump at a continuous Mott transition between a FL and a spin liquid~\cite{TheoryContinuousMottsenthil2008a,UniversalTransportQuantumwitczak-krempa2012}.
The jump is a direct consequence of fractionalization.
{The internal gauge constraint enforces a series composition of
the charged- and neutral-sector resistivities, known as the
Ioffe--Larkin (IL) composition rule~\cite{GaplessFermionsGaugeioffe1989,PhysRevB.46.5621}.
At $\nu=-1/2$, this directly gives the measured resistivity.
At $\nu=-3/4$, it gives the quarter-filled reference response,
from which the measured tensor is obtained by particle--hole
reconstruction. The spinon Fermi surface supplies a smooth internal
metallic contribution, while the condensed boson contributes no
resistance in the FL. We quote the quantum critical point (QCP)--FL
resistivity jumps in the clean spinon limit. At $\nu=-1/2$, the jump is the
critical charged-sector tensor itself, whereas at $\nu=-3/4$
it follows from that tensor after particle--hole reconstruction.}

Computing this universal contribution poses four difficulties.
First, the Chern--Simons terms mix the emergent gauge fields with one another and with the external probe field.
The composition rule for the full tensor, including its Hall components, therefore differs from the scalar series addition of parton theories without Chern--Simons terms and must be derived for each critical theory.
Second, the limits of zero probe frequency $\omega$ and zero temperature $T$ do not commute at a quantum critical point~\cite{NonzerotemperatureTransportQuantumdamle1997,NonzerotemperatureTransportFractionalsachdev1998a}.
A zero-temperature Kubo calculation accesses only the collisionless regime $\hbar\omega\gg k_BT$, whereas the DC response belongs to the hydrodynamic regime $\hbar\omega\ll k_BT$ and is controlled by collisions of thermally excited carriers with the gauge fluctuations.
A calculation without collisions yields a Drude peak of zero width and hence no finite DC conductivity.
Third, the critical theory is strongly coupled and has no small parameter.
Fourth, Landau damping by the spinon Fermi surface can generate, at large spinon Fermi energy, a zero-frequency singularity in the Bose-weighted gauge spectrum: it occupies a vanishingly narrow frequency window yet carries finite scattering weight, and would act as an additional quasistatic scattering channel at nonzero temperature~\cite{song_2024phase,UniversalTransportQuantumwitczak-krempa2012}.

We address these difficulties in two steps: the Gaussian solution of the gauge sector resolves the first, and the large-$N$ quantum Boltzmann equation resolves the remaining three.
We first integrate out the matter fields to quadratic order and solve the resulting Gaussian gauge theory exactly.
This yields the IL composition rule for the full resistivity tensor of each critical theory and reduces the problem to a single sector quantity, the longitudinal conductivity $\sigma_\psi$ of the critical Dirac fermions, defined as their current response to the electric field of the emergent gauge field they couple to.
This step requires keeping track of the integer Chern--Simons contact term fixed by the regularization of the Dirac sector, which at the critical point cancels the explicit Chern--Simons term of the gauge field coupled to the Dirac fermions (Sec.~\ref{sec:crithe}).
We then compute $\sigma_\psi$ in a large-$N$ expansion, in which the complete matter content is replicated $N$ times and the Chern--Simons levels are scaled with $N$ so that Chern--Simons mixing remains of the same order as the matter polarization~\cite{QuantumCriticalityU1kaul2008,PhysRevB.82.045121}.
The Dirac carriers are collisionless at $N=\infty$ and scatter off emergent gauge fluctuations at order $1/N$.
We treat these collisions with a quantum Boltzmann equation (QBE) for thermally excited carriers in the positive- and negative-energy Dirac bands~\cite{NonzerotemperatureTransportQuantumdamle1997,NonzerotemperatureTransportFractionalsachdev1998a,QuantumCriticalTransportfritz2008a,PhysRevB.52.17275}.
Building the collision kernel from the Coulomb-gauge propagator at finite spinon Fermi energy, we find that this zero-frequency structure survives the Chern--Simons mixing: the Bose-weighted gauge spectrum develops, in addition to its finite-frequency part, a quasistatic peak of width $\sim T^2/\mu$ and height $\sim\mu/T^2$, whose integrated weight is finite and independent of both $\mu$ and the spinon Fermi velocity. The density and transverse diagonal channels both contribute. The collision operator therefore contains, in addition to inelastic scattering from the finite-frequency gauge spectrum, a quasistatic elastic channel, which we include explicitly.
Solving the QBE yields a universal scaling function of $\omega/T$ whose zero-frequency limit gives a finite DC conductivity.
Following earlier large-$N$ studies of quantum critical transport~\cite{UniversalTransportQuantumwitczak-krempa2012,QuantumCriticalTransportfritz2011}, we evaluate this leading-order result at the physical value $N=1$.

{Our results are summarized by the charged-sector composition rule
\begin{equation}
\label{eq:intro_IL}
\rho_\Phi
=
n_\psi\left[
\sigma_\psi^{-1}\mathbb{I}
+\sigma_{\mathrm{CS}}^{-1}
\right].
\end{equation}}

Here $\mathbb I$ is the $2\times2$ identity,
$\sigma_{\mathrm{CS}}=(e^2/h)\epsilon$ with $\epsilon_{xy}=+1$,
and $n_\psi=1$ $(2)$ is the number of two-cone Dirac sectors
at $\nu=-1/2$ $(-3/4)$.
The quantity $\sigma_\psi$ is the dynamical longitudinal
conductivity of one such sector, with the fixed Chern--Simons
contact term removed.
At $\nu=-1/2$, $\rho=\rho_f\mathbb I+\rho_\Phi$.
At $\nu=-3/4$, this series combination is the quarter-filled
reference response; the measured DC tensor is reconstructed
according to Eq.~\eqref{eq:rho_phys_iso}.
The quoted universal quantities are the clean-limit,
low-temperature DC differences $\Delta\rho$ between the QCP
and the adjacent FL, defined in Eq.~\eqref{eq:r_definition}.
In terms of $g_\psi=(h/e^2)\sigma_\psi(0)$ they read
\begin{equation}
\label{eq:intro_jump}
(\Delta\rho_{xx},\,\Delta\rho_{xy})
=
\begin{cases}
\left(\dfrac{1}{g_\psi},\,-1\right)\dfrac{h}{e^2},
& \nu=-1/2,\\[8pt]
\left(\dfrac{2g_\psi}{4+g_\psi^2},\,
-\dfrac{4+2g_\psi^2}{4+g_\psi^2}\right)\dfrac{h}{e^2},
& \nu=-3/4.
\end{cases}
\end{equation}
Using the leading-order large-$N$ estimates evaluated at
$N=1$, we obtain the values collected in Table~\ref{tab:summary}.

\begin{table}[h]
\centering
\begin{tabular}{@{}ccccc@{}}
\hline
$\nu$ & $n_\psi$ & $\sigma_\psi(0)\,[e^2/h]$ & $\Delta\rho_{xx}\,[h/e^2]$ & $\Delta\rho_{xy}\,[h/e^2]$\\
\hline
$-1/2$ & 1 & 0.1696 & 5.896 & $-1$\\
$-3/4$ & 2 & 0.2591 & 0.1274 & $-1.0165$\\
\hline
\end{tabular}
\caption{Leading-order large-$N$ estimates of the Dirac-sector DC conductivity and the clean-limit resistivity jumps at $N=1$.}
\label{tab:summary}
\end{table}

\paragraph{Conventions}
We use $\hbar=k_B=e=1$, a Dirac velocity $v_\psi=1$, and therefore $e^2/h=1/(2\pi)$; conductivities are computed in units of $e^2/\hbar$ and reported in units of $e^2/h$, with resistivities in $h/e^2$. The spinon parameters $\mu$, $m$, and $v_F=\sqrt{2\mu/m}$ are kept explicit. Euclidean gauge and fermion three-momenta are $P=(\Omega_n,\boldsymbol q)$ and $K=(\omega_n,\boldsymbol k)$; the scalar $P$ is the Euclidean magnitude, continued to $P=\sqrt{q^2-(\Omega+i0^+)^2}$ at real frequency. The external probe frequency is $\omega$, while $\Omega$ denotes the gauge-boson energy transfer. We use $\epsilon^{012}=\epsilon_{xy}=+1$, denote the measured resistivity tensor by $\rho$, and write $\sigma^{\rm phys}=\rho^{-1}$; the charged-sector resistivity is $\rho_\Phi$.

The rest of the paper is organized as follows.
Section~\ref{sec:crithe} reviews the critical theory at $\nu=-1/2$, and Sec.~\ref{sec:IL} derives the IL composition rule.
Section~\ref{sec:largeN} introduces the large-$N$ expansion, analyzes the collisionless conductivity, and derives the gauge propagator and spectral functions. Section~\ref{sec:qbe} solves the QBE for $\nu=-1/2$, Sec.~\ref{sec:34} extends the analysis to $\nu=-3/4$, and Sec.~\ref{sec:discussion} discusses high-frequency and thermal transport together with experimental implications.

\section{Critical theory of the CFL--FL transition at $\nu=-1/2$}
\label{sec:crithe}

In this section we review the critical theory for the $\nu=-1/2$ CFL--FL transition and summarize the expected scaling structure of resistivity on the zero-detuning critical trajectory.

Motivated by recent observations of fractional quantum anomalous Hall (FQAH) states and composite Fermi liquid (CFL) phenomenology in moir\'e Chern bands, Song, Zhang and Senthil~\cite{song_2024phase} proposed effective field theories for bandwidth-tuned phase transitions out of quantum Hall states at fixed lattice filling. A central idea is a parton description in which the physical electron fractionalizes into a charged sector and a neutral Fermi-surface sector, coupled by an emergent $U(1)$ gauge field. In this framework, quantum Hall/CFL physics is encoded in the charged sector, while gapless metallic degrees of freedom can persist in a neutral sector; their coupling through gauge fluctuations enables a unified description of the CFL--FL transition.

{We focus on the transition at filling $\nu=-1/2$,
using a direct half-filled $C=1$ carrier description.}

Following Ref.~\cite{song_2024phase}, we write the low-energy electron operator $c$ (in a single, valley-polarized $C=1$ moir\'e Chern band) in the parton form
\begin{equation}
    c = \Phi f,
\end{equation}
where $\Phi$ is a charge-$1$ boson and $f$ is an electrically neutral fermion.
This introduces an emergent internal $U(1)$ gauge redundancy $\Phi\to e^{i\theta}\Phi$, $f\to e^{-i\theta}f$, enforced by a dynamical gauge field $a_\mu=(a_0,\boldsymbol a)$.
The external probe gauge field $A_\mu$ couples to the conserved physical electric charge; with the above assignment, $\Phi$ couples to $A_\mu+a_\mu$, while $f$ couples to $-a_\mu$ and forms a neutral Fermi surface.

At $\nu=-1/2$, the CFL--FL transition corresponds to driving the charged boson $\Phi$ between a bosonic $\nu_\Phi=1/2$ Laughlin state (CFL) and a superfluid (FL).
A convenient critical description further fractionalizes the boson as
\begin{equation}
    \Phi = f_1 f_2,
\end{equation}
which introduces another emergent gauge field (denoted $b_\mu$ below).
In the mean-field construction of Ref.~\cite{song_2024phase}, tuning the bandwidth drives a Chern-number-changing transition in one fermionic parton sector, yielding a continuum Dirac theory at low energies.
The resulting critical Lagrangian (in real time) can be written as
\begin{equation}
    \mathcal{L}_{\mathrm{crit}}
    = \mathcal{L}_{\psi}+\mathcal{L}_{f}+\mathcal{L}_{\mathrm{CS}},
\end{equation}
with
\begin{equation}
\begin{aligned}
    \mathcal{L}_{\psi}
    &= -i\bar{\psi}\left(\slashed{\partial}-i\slashed{b}\right)\psi
       + M\bar{\psi}\psi,\\
    \mathcal{L}_{f}
    &= f^\dagger\left[
       i\partial_t-a_0+\mu
       +\frac{1}{2m}(\nabla+i\boldsymbol a)^2
       \right]f,\\
    \mathcal{L}_{\mathrm{CS}}
    &= \frac{1}{4\pi}(b+a+A)\,\mathrm d(b+a+A)
       -\frac{1}{4\pi}b\,\mathrm db.
\end{aligned}
\end{equation}

Here $\psi$ is the critical Dirac fermion coupled minimally to the emergent gauge field $b_\mu$, while $f$ forms a neutral Fermi surface coupled to $a_\mu$.
The field $\psi$ contains two physical two-component Dirac cones. Their flavor index is suppressed, and bilinears are summed over this physical flavor space.
The Chern--Simons terms encode the topological response of the charged sector and mix the internal gauge fields with the probe field.
In particular, the term of the form $\frac{1}{2\pi}b\,\mathrm d(a+A)$ that appears upon expanding $(b+a+A)\mathrm d(b+a+A)$ is a \emph{BF coupling} (a mutual Chern--Simons term): it couples the flux of one gauge field to the charge of the other and transmits the physical $U(1)$ response through gauge-invariant combinations.

The Dirac term in $\mathcal L_\psi$ must be supplemented by a regularization prescription, which fixes the parity-odd contact term of the two-cone response. We write the full response of the regularized Dirac sector to $b$ as
\begin{equation}
\label{eq:contact}
\widehat\Pi_\psi=\Pi_\psi+n_{\mathrm{ct}}\,\Pi_{\mathrm{CS}},
\end{equation}
where $\Pi_{\mathrm{CS}}$ is the unit-level Chern--Simons kernel [Eq.~\eqref{eq:PiCS_def}], $\Pi_\psi$ is the dynamical part, and $n_{\mathrm{ct}}$ is an integer fixed by the Chern numbers of the lattice parton bands on the two sides of the transition. On the two massive-Dirac sides the two cones give $\widehat\Pi_\psi=(n_{\mathrm{ct}}\pm1)\Pi_{\mathrm{CS}}$. Including the explicit term $-\frac{1}{4\pi}b\,\mathrm db$, the coefficient of
$b\,\mathrm db$ outside $(b+a+A)\,\mathrm d(b+a+A)$ is $n_{\mathrm{ct}}-1\pm1$.
The total $b\,\mathrm db$ level is $n_{\mathrm{ct}}\pm1$. For $n_{\mathrm{ct}}=1$ the two levels are $0$ and $2$: the superfluid side is the $0$ level, and the bosonic Laughlin side is the $2$ level. As a direct check, setting $X=a+A$ and taking the total level $2$, integrating out $b$ gives $X\,\mathrm dX/(8\pi)$, the bosonic $\nu_\Phi=1/2$ Laughlin response. Hence
\begin{equation}
\label{eq:kappa}
n_{\mathrm{ct}}=1,\qquad
\widehat\Pi_\psi-\Pi_{\mathrm{CS}}=\Pi_\psi .
\end{equation}
At the critical point the $+1$ contact term cancels the explicit $-\frac{1}{4\pi}b\,\mathrm db$; the shift term remains, and $\Pi_\psi$ is parity even at leading order. A different regularization convention moves this integer between the Dirac term and the bare Chern--Simons level, but their sum is convention independent. In the large-$N$ theory of Sec.~\ref{sec:largeN} each replica carries the same contact term, so it scales with $N$ like the other Chern--Simons levels.

The detuning $\delta$ tunes the effective Dirac mass $M$ through zero. Changing the sign of $M$ changes the induced Chern--Simons response of the $\psi$ sector, so that after integrating out gapped modes one obtains either (i) the bosonic Laughlin phase for $\Phi$ (corresponding to the electronic CFL) or (ii) a condensed $\Phi$ superfluid (corresponding to the electronic FL via Higgsing of $a_\mu$)~\cite{song_2024phase}.
The bosonic parton $\Phi$ is not a simple bilinear of $\psi$; rather, it is represented by a $b$-gauge-invariant monopole operator in the critical theory, carrying unit charge under both $a_\mu$ and $A_\mu$ through the BF coupling.

As a scale-invariant critical point, the theory implies universal scaling forms for observables in the quantum critical regime.
{In particular, the electrical resistivity tensor obeys
\begin{equation}
    \rho_{ij}-\rho_f\delta_{ij}
    =\frac{h}{e^2}
    \mathcal{R}_{ij}\!\left(
    \frac{\omega}{T},
    \frac{\delta}{T^{1/\nu_{\rm corr}z}}
    \right).
\end{equation}
Here $\delta$ is the tuning parameter across the transition,
$\nu_{\mathrm{corr}},z$ are the correlation-length and dynamical exponents,
and $\rho_f$ is the smooth spinon resistivity.
This form applies to the direct half-filled response considered
in this section.}

In the following section, we derive the Ioffe--Larkin composition rule for $\rho_{ij}$ by integrating out the matter fields and obtaining an effective Gaussian gauge theory. Solving this theory allows us to reduce the transport problem to computing the universal contribution from the critical $\psi$ sector and then performing the required Chern--Simons composition.

\begin{figure}
    \centering
    \includegraphics[width=\linewidth]{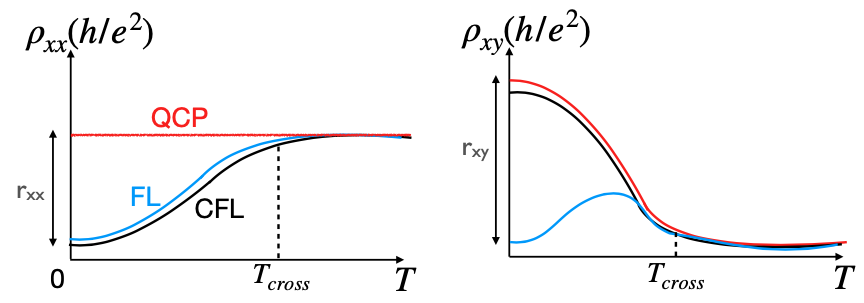}
    \caption{Qualitative sketch of the temperature dependence of the resistivities,
    adapted from Fig.~5 of Ref.~\cite{song_2024phase}.
    The blue, black, and red curves denote the FL, CFL, and QCP,
    respectively. The arrows $r_{xx}$ and $r_{xy}$ indicate the
    QCP--FL separations in the sketch.}
    \label{fig:resi}
\end{figure}

Figure~\ref{fig:resi} illustrates the temperature crossover sketched in
Ref.~\cite{song_2024phase}. The clean-limit DC jumps considered here
are defined in Eq.~\eqref{eq:r_definition}.

\section{Ioffe--Larkin composition rule and the resistivity tensor}
\label{sec:IL}

In this section we derive the Ioffe--Larkin (IL) composition rule for the \emph{full} $2\times 2$ electrical resistivity tensor of the physical electrons. The strategy is standard: we first integrate out the gapless matter fields (the neutral Fermi surface $f$ and the critical Dirac fermions $\psi$), which generates polarization tensors for the emergent gauge fields. The remaining theory of gauge fluctuations is then quadratic (Gaussian), so the exact linear electromagnetic response within this quadratic theory follows from solving the saddle-point equations for the internal gauge fields and substituting back.

We start from the Euclidean continuation of $\mathcal{L}_{\mathrm{crit}}$,
\begin{equation}
\mathcal{L}_{\mathrm{crit}}
=\mathcal{L}_\psi[\psi,b]
+\mathcal{L}_f[f,a]
+\mathcal{L}_{\mathrm{CS}}[a,b,A],
\end{equation}
and integrate out the fermions $\psi$ and $f$ at fixed background gauge fields $b_\mu$ and $a_\mu$. This defines the (Euclidean) polarization tensors
\begin{align}
    \widehat\Pi^{\mu\nu}_\psi(P)
    \equiv&
    \left.
    \frac{\delta^2\ln Z_\psi[b]}
    {\delta b_\mu(-P)\,\delta b_\nu(P)}
    \right|_{b=0},\\
    \Pi^{\mu\nu}_f(P)
    \equiv&
    \left.
    \frac{\delta^2\ln Z_f[a]}
    {\delta a_\mu(-P)\,\delta a_\nu(P)}
    \right|_{a=0},
\end{align}
with $P\equiv(\omega,\boldsymbol p)$.
At small $|\boldsymbol p|,\;|\omega|$, gauge invariance implies transversality
$P_\mu\Pi^{\mu\nu}=0$ for both $\widehat\Pi_\psi$ and $\Pi_f$.
In the $p\to0$ transport limit, $\Pi^{ij}$ encodes the optical conductivity via the Kubo relation
$\sigma^{ij}(\omega)=\lim_{\boldsymbol p\to0}\Pi^{ij}(\omega,\boldsymbol p)/i\omega$.

Throughout this paper, $\Pi$ denotes the response kernel
defined by the second derivative of $\ln Z$, rather than
the quadratic kernel of $-\ln Z$.
Consequently, integrating out matter contributes
$-\frac12\!\int\! x_\mu^*\Pi^{\mu\nu}x_\nu$
to the Euclidean effective action.
Retarded kernels are obtained by analytic continuation
$i\omega_n\to\omega+i0^+$.
We use $\epsilon^{012}=+1$ and $\epsilon_{xy}=+1$.

The Euclidean Chern--Simons response kernel is
\begin{equation}
\label{eq:PiCS_def}
\Pi_{\mathrm{CS}}^{\mu\nu}(P)
=
-\frac{1}{2\pi}\epsilon^{\mu\nu\rho}P_\rho.
\end{equation}
In the homogeneous retarded transport limit, this gives
\begin{equation}
\Pi_{\mathrm{CS}}^{ij,R}(\omega,\boldsymbol0)
=
\frac{i\omega}{2\pi}\epsilon^{ij}.
\end{equation}
With the response-kernel convention used here, a
Euclidean term $-\frac{i}{4\pi}\!\int\! x\,\mathrm dx$
contributes
$-\frac12\!\int\! x_\mu^*\Pi_{\mathrm{CS}}^{\mu\nu}x_\nu$.

The full Dirac response contains the contact term of Eqs.~\eqref{eq:contact} and \eqref{eq:kappa}, $\widehat\Pi_\psi=\Pi_\psi+\Pi_{\mathrm{CS}}$. From here on $\Pi_\psi$ denotes the dynamical part, and $\sigma_\psi$ the corresponding longitudinal conductivity. Integrating out $\psi$ contributes $-\frac12\!\int\! b_\mu^*\widehat\Pi_\psi^{\mu\nu}b_\nu$; its contact term cancels the contribution $+\frac12\!\int\! b_\mu^*\Pi_{\mathrm{CS}}^{\mu\nu}b_\nu$ of the explicit $-\frac{1}{4\pi}b\,\mathrm db$.

After integrating out $\psi$ and $f$, the effective quadratic Euclidean action for $(a,b)$ in the presence of $A$ becomes
\begin{equation}
\label{eq:SE_quad}
\begin{aligned}
\mathcal{S}_E[b,a;A]
=-\frac12\!\!\int_{\omega,\boldsymbol p}\!
\Big[
b_\mu^*\Pi^{\mu\nu}_\psi b_\nu
+a_\mu^*\Pi^{\mu\nu}_f a_\nu +(b_\mu+a_\mu+A_\mu)^*
\Pi^{\mu\nu}_{\mathrm{CS}}
(b_\nu+a_\nu+A_\nu)
\Big],
\end{aligned}
\end{equation}
where the Fourier components obey
$x(\boldsymbol p,\omega)=x^*(-\boldsymbol p,-\omega)$.

The probe field $A$ appears only through the combination $(b+a+A)$ in the Chern--Simons term.
Consequently, the physical electromagnetic current extracted from the generating functional,
\begin{equation}
\label{eq:Jphy_def}
J_\mu^{\mathrm{phys}}(P)
\equiv
-\frac{\delta\mathcal{S}_E}{\delta A_\mu(-P)},
\end{equation}
is proportional to the dual field strength of $b+a+A$, because the microscopic charge is carried by the $\Phi$ sector and is transmitted to the infrared (IR) through the topological CS/BF structure.

Evaluating Eq.~\eqref{eq:Jphy_def} using Eq.~\eqref{eq:SE_quad} gives
\begin{equation}
\label{eq:Jphy_baA}
J_\mu^{\mathrm{phys}}(P)
=\Pi^{\mu\nu}_{\mathrm{CS}}(P)
\big[b_\nu(P)+a_\nu(P)+A_\nu(P)\big].
\end{equation}
After solving for $a,b$ in terms of $A$, this becomes
$J^{\mathrm{phys}}_\mu=\Pi^{\mu\nu}_{\mathrm{phys}}A_\nu$,
defining the physical polarization tensor $\Pi^{\mathrm{phys}}$.

Because Eq.~\eqref{eq:SE_quad} is quadratic, the response follows from the saddle-point equations
\begin{equation}
\label{eq:eom_ab}
\left\{
\begin{aligned}
\frac{\delta\mathcal{S}_E}{\delta b_\mu^*}=0:
&\quad
\Pi^{\mu\nu}_\psi b_\nu
+\Pi^{\mu\nu}_{\mathrm{CS}}(b_\nu+a_\nu+A_\nu)=0,\\
\frac{\delta\mathcal{S}_E}{\delta a_\mu^*}=0:
&\quad
\Pi^{\mu\nu}_f a_\nu
+\Pi^{\mu\nu}_{\mathrm{CS}}(b_\nu+a_\nu+A_\nu)=0.
\end{aligned}
\right.
\end{equation}
Solving these linear equations for $(a,b)$ and substituting into Eq.~\eqref{eq:Jphy_baA} yields an IL-type composition rule for the physical polarization tensor:
\begin{equation}
\label{eq:IL_Pi}
(\Pi^{\mathrm{phys}})^{-1}
=\Pi_{\mathrm{CS}}^{-1}
+\Pi_f^{-1}
+\Pi_\psi^{-1}.
\end{equation}
In terms of the full Dirac response, the last term is $(\widehat\Pi_\psi-\Pi_{\mathrm{CS}})^{-1}$.
The inverse structure reflects that, once gauge constraints are imposed, the corresponding response channels combine as impedances in series.

In the homogeneous transport limit $\boldsymbol p\to0$, gauge invariance enforces
$\Pi^{0i}(\omega,\boldsymbol0)=0$ for $\omega\neq0$, so the spatial $2\times2$ block fully determines charge transport:
\begin{equation}
\sigma^{ij}(\omega)
=\frac{1}{i\omega}\Pi^{ij}(\omega,\boldsymbol0),
\qquad
\rho(\omega)=\sigma(\omega)^{-1}.
\end{equation}
{In units with $e=\hbar=1$, the unit-level Chern--Simons
response is
$\sigma_{\mathrm{CS}}=\epsilon/(2\pi)$ and
$\sigma_{\mathrm{CS}}^{-1}=-2\pi\epsilon$.
Equation~\eqref{eq:IL_Pi} therefore gives
\begin{equation}
\label{eq:rho_IL_tensor}
\begin{aligned}
\rho&=\rho_f\mathbb I+\rho_\Phi,\\
\rho_\Phi&=
\sigma_\psi^{-1}\mathbb I
+\sigma_{\mathrm{CS}}^{-1}.
\end{aligned}
\end{equation}
Here $\sigma_\psi$ is the dynamical longitudinal conductivity
of the critical two-cone Dirac sector, and the spinon
response is taken to be the isotropic scalar $\rho_f$.
Explicitly,
\begin{equation}
\label{eq:rho_explicit}
(\rho_\Phi)_{ij}
=
\frac{1}{\sigma_\psi}\delta_{ij}
-2\pi\epsilon_{ij}.
\end{equation}
At leading order, the longitudinal part is the inverse Dirac conductivity,
and the Hall part is fixed by the Chern--Simons level.
If the dynamical Dirac response acquires a Hall component at higher orders,
$\sigma_\psi^{-1}\mathbb I$ is replaced by the inverse of the
full dynamical conductivity tensor.

Across the transition, $\rho_f$ supplies a smooth internal
metallic contribution, while the critical charged-sector
tensor is determined by $\sigma_\psi$.
For comparison with the adjacent FL, we define
\begin{equation}
\label{eq:r_definition}
\begin{aligned}
\Delta\rho_{ij}
={}&
\lim_{T\to0}\lim_{\omega\to0}
\rho_{ij}(0,T,\omega)\\
&-
\lim_{\delta\to0\,({\mathrm{FL}})}
\lim_{T\to0}\lim_{\omega\to0}
\rho_{ij}(\delta,T,\omega).
\end{aligned}
\end{equation}
The quoted values use the clean spinon limit,
$\rho_f\to0$.
The DC limit is taken at nonzero temperature before the
low-temperature limit; the FL limit is taken from the FL side
after its low-temperature response is obtained.
Thus $\Delta\rho$ compares the QCP with the adjacent FL,
not the two metallic phases at arbitrary detuning.
At $\nu=-1/2$, the clean-limit jump equals the critical
$\rho_\Phi$.
The measured $\nu=-3/4$ tensor must first be reconstructed
as described in Sec.~\ref{sec:experiment}.}

Microscopically, $\sigma_\psi$ is controlled by the low-energy
dynamics of the gauge field $b$ to which $\psi$ couples.
We therefore determine the infrared effective theory of $b$
and compute $\sigma_\psi$ using a large-$N$ expansion and
quantum Boltzmann equation before performing the
Chern--Simons composition.

\section{Large-$N$ formulation}\label{sec:largeN}

\subsection{Controlled large-$N$ construction}

To extract the leading correction to the fermion conductivity, we employ a large-$N$ expansion. We replicate the physical matter content \(N\) times and simultaneously promote the levels of the Chern--Simons terms to \(N\). This choice keeps the Chern--Simons contributions at the same parametric order as the matter-induced polarization effects, so that their impact remains explicit throughout the $1/N$ expansion. Our large-\(N\) construction is in the same spirit as controlled expansions developed for other non-Fermi-liquid metals with singular bosonic interactions~\cite{PhysRevB.82.045121}, although here the presence of Chern--Simons terms requires a different scaling assignment.

Each replica $\psi_l$ contains the two physical Dirac cones, so $N=1$ recovers the physical theory, and the Chern--Simons levels divided by $N$ are held fixed.

The corresponding imaginary-time Lagrangian in the absence of the external field $A$ is
\begin{equation}
\mathcal{L}_N
=
\sum_{l=1}^N
\left(
\mathcal{L}_{\psi,l}+\mathcal{L}_{f,l}
\right)
-\frac{iN}{4\pi}(a+b)\,\mathrm d(a+b),
\end{equation}
with
\begin{align}
\mathcal{L}_{\psi,l}
&=
-\overline{\psi_l}
(\slashed{\partial}-i\slashed{b})\psi_l
-M\overline{\psi_l}\psi_l,\\
\mathcal{L}_{f,l}
&=
f_l^\dagger
\left[
\partial_\tau+ia_0-\mu
-\frac{1}{2m}(\nabla+i\boldsymbol a)^2
\right]f_l.
\end{align}
The Dirac bubble below is the dynamical response $\Pi_\psi$.

We adopt the Euclidean $\gamma$ matrices
$\gamma^0=\sigma^z$, $\gamma^1=-\sigma^y$, and $\gamma^2=\sigma^x$,
so that
$\bar\psi\gamma^i\psi=i\psi^\dagger\sigma^i\psi$.

Working in the Lorenz gauge, the free propagators can be directly read off from $\mathcal{L}_N$:

\begin{figure}[t]
    \centering
    \subfloat[Fermion bubble diagram]{
    \begin{tikzpicture}[baseline=(current bounding box.center)]
    \begin{feynman}
      \vertex (a);
      \vertex [right=0.5 cm of a] (b);
      \vertex [right=0.8 cm of b] (c);
      \vertex [right=0.5 cm of c] (d);
      \diagram* {
        (a) -- [photon] (b),
        (b) -- [fermion, half left] (c),
        (c) -- [fermion, half left] (b),
        (c) -- [photon] (d)
      };
      \path[use as bounding box]
      ([shift={(-0.4,-0.4)}]current bounding box.south west)
      rectangle
      ([shift={(0.4,0.4)}]current bounding box.north east);
    \end{feynman}
    \end{tikzpicture}
    \label{fig:bubble}
    }
    \quad
    \subfloat[Leading correction of fermion self-energy]{
    \begin{tikzpicture}[baseline=(current bounding box.center)]
    \begin{feynman}
      \vertex (a);
      \vertex [right=0.5 cm of a] (b);
      \vertex [right=0.8 cm of b] (c);
      \vertex [right=0.5 cm of c] (d);
      \diagram* {
        (a) -- [fermion] (b),
        (b) -- [photon, half left] (c),
        (b) -- [fermion] (c),
        (c) -- [fermion] (d)
      };
      \path[use as bounding box]
      ([shift={(-0.4,-0.4)}]current bounding box.south west)
      rectangle
      ([shift={(0.4,0.4)}]current bounding box.north east);
    \end{feynman}
    \end{tikzpicture}
    \label{fig:self-energy}
    }
    \caption{Leading one-loop diagrams. Solid lines represent free propagators of the Dirac fermion $\psi$, and wavy lines represent propagators of the gauge field $b$. Diagram (a) is of order $N$ since each replica can form a loop, while diagram (b) is of order $N^{-1}$ and is thus suppressed.}
    \label{fig:feynman}
\end{figure}

\begin{align}
\Delta_a^{(0)}(P)
&=
\left[
-N(\Pi_{\mathrm{CS}})_{\mu\nu}(P)
+\frac{N}{\xi}P_\mu P_\nu
\right]^{-1},\\
G_\psi(P)
&=
\frac{1}{i\slashed P+M},\\
G_f(P)
&=
\left(
-i\omega_n+\frac{\boldsymbol p^2}{2m}-\mu
\right)^{-1}.
\end{align}
Here $\xi$ is an infinitesimal positive gauge-fixing parameter.

In the large-$N$ theory, perturbation theory is organized into a controlled expansion in powers of $1/N$. Each closed matter loop carries a factor of $N$ from the sum over replicas, while the gauge sector remains of the same parametric order because the Chern--Simons terms have also been scaled by $N$. As a result, the dominant one-particle-irreducible contribution to the gauge polarization is the fermion bubble in Fig.~\ref{fig:bubble}, whereas higher-order corrections are suppressed by additional powers of $1/N$.

For fermion self-energy diagrams, the leading correction is of $N^{-1}$ order, as shown in Fig.~\ref{fig:self-energy}. We therefore neglect the mass renormalization at leading order and set $M=0$ at the critical point within the large-$N$ treatment.

The one-loop bubble diagrams of $\psi$ and $f$ read:

\begin{align}
&\Pi_\psi^{\mu\nu}
\left(P\right)
=
-2T\sum_{\omega_n}
\!\int\!\frac{\mathrm d^2\boldsymbol k}{(2\pi)^2}
\frac{
2\big\{
2K^{(\mu}(K+P)^{\nu)}
-
[\boldsymbol k\!\cdot\!(\boldsymbol k+\boldsymbol q)
+\omega_n(\omega_n+\Omega_n)]
\delta^{\mu\nu}
\big\}
}{
(\varepsilon_{\boldsymbol k}^2+\omega_n^2)
[\varepsilon_{\boldsymbol k+\boldsymbol q}^2+(\omega_n+\Omega_n)^2]
},
\label{eq:pi_psi}\\
&\Pi_f^{ij}(P)
=
T\sum_{\omega_n}
\!\int\!\frac{\mathrm d^2\boldsymbol k}{(2\pi)^2}
\left(
\frac{\delta^{ij}}{m}G_f(K)
-
\frac{1}{m^2}
\left(k-q/2\right)^i
\left(k-q/2\right)^j
G_f(K)G_f(K-P)
\right),
\label{eq:pi_fii}\\
&\Pi_f^{00}(P)
=
T\sum_{\omega_n}
\!\int\!\frac{\mathrm d^2\boldsymbol k}{(2\pi)^2}
G_f(K)G_f(K-P).
\label{eq:pi_f00}
\end{align}

Here, $P\equiv(\Omega_n,\boldsymbol q)$. These leading polarization functions are the basic input for the collisionless conductivity below and for the infrared effective theory of the gauge field $b$ in Sec.~\ref{sec:gauge}. Explicit retarded forms are given in Appendix~\ref{app:polarization}.

\subsection{Collisionless conductivity at $N=\infty$}

We first consider the conductivity of the \(\psi\) sector in the collisionless limit. At leading order in the large-\(N\) expansion, \(\sigma_\psi\) is determined by the Dirac-fermion polarization bubble,
\begin{align}
\sigma_\psi(i\Omega_n)
&=
-\frac{1}{\Omega_n}
\Pi_\psi^{xx}(i\Omega_n,\boldsymbol 0)
\nonumber\\
&=
-\frac{T}{\Omega_n}
\sum_{\omega_n}
\!\int\!\frac{\mathrm d^2\boldsymbol k}{(2\pi)^2}
\frac{
4\omega_n(\omega_n+\Omega_n)
}{
(\varepsilon_{\boldsymbol k}^2+\omega_n^2)
[\varepsilon_{\boldsymbol k}^2+(\omega_n+\Omega_n)^2]
}.
\label{eq:sigma_psi_bubble}
\end{align}
This result is well known from studies of finite-temperature transport in relativistic quantum critical theories~\cite{NonzerotemperatureTransportFractionalsachdev1998a}. At order \(N^0\), the conductivity decomposes into
\begin{align}
\sigma_\psi(\omega)
&=
\sigma_{\mathrm{qp}}(\omega)
+\sigma_{\mathrm{coh}}(\omega),\\
\sigma_{\mathrm{qp}}(\omega)
&=
\ln2\,\delta\!\left(\frac{\omega}{T}\right)
+
i\,\frac{\ln2}{\pi}
\mathcal P\!\left(\frac{T}{\omega}\right),
\label{eq:sharp}\\
\operatorname{Re}\,\sigma_{\mathrm{coh}}(\omega)
&=
\frac18
\tanh\!\left(\frac{|\omega|}{4T}\right).
\label{eq:re_coh}
\end{align}
The term \(\sigma_{\mathrm{qp}}\) originates from Dirac quasiparticles thermally populated in the positive- and negative-energy bands and yields a singular Drude response in the collisionless limit. The term \(\sigma_{\mathrm{coh}}\) comes from the direct excitation of particle--hole pairs by the probe field. It is therefore the intrinsic collisionless interband contribution and is already finite in the free-bubble approximation.

These two terms dominate in different frequency regimes. For $\omega\gg T$, thermal occupation is negligible and the conductivity is controlled by $\sigma_{\mathrm{coh}}$, whose real part approaches the constant value
\begin{equation}
\label{eq:inf_limit}
\operatorname{Re}\,\sigma_\psi(\omega)
\to
{\frac{\pi}{4}\frac{e^2}{h}}.
\end{equation}
This is the leading-order collisionless value for a relativistic Dirac theory. By contrast, in the low-frequency regime $\omega\lesssim T$, thermally excited carriers become important, and their contribution is encoded in $\sigma_{\mathrm{qp}}$. The $\delta(\omega/T)$ term in Eq.~\eqref{eq:sharp} is a Drude peak with zero width, while the principal-value imaginary part is its Kramers--Kronig partner.

At $N=\infty$ the thermally excited carriers are accelerated by the probe field but never relax, which produces the zero-width Drude peak in Eq.~\eqref{eq:sharp}. Gauge-mediated collisions at order $1/N$ broaden this peak. Capturing them in the low-frequency limit requires resumming an infinite class of scattering processes, which we do with a quantum Boltzmann equation in Sec.~\ref{sec:qbe}.

\subsection{Gauge propagator and spectral functions}
\label{sec:gauge}

The collision integral of the quantum Boltzmann equation is built from the propagator of the emergent gauge field \(b\), which we now construct. Integrating out all other fields gives an effective action quadratic in \(b\),
\begin{equation}
S_{\text{eff}}[b]
=
-\frac{N}{2}
\!\int\!\frac{\mathrm d^3P}{(2\pi)^3}
b_\mu(-P)\!\cdot\!\Pi_b^{\mu\nu}(P)\!\cdot\! b_\nu(P).
\end{equation}
The response kernel $\Pi_b$ is defined so that the
Euclidean propagator satisfies
$\Delta_b^{-1}=-N\Pi_b$. In the large-$N$ expansion, $\Pi_b$ is determined by the leading matter polarization tensors $\Pi_\psi$ and $\Pi_f$, together with the Chern--Simons mixing between the gauge fields.

To compute $\Pi_b$, we first write the inverse propagator relations implied by the Feynman rules:
\begin{equation}
\left\{
\begin{aligned}
&\Delta_a^{-1}
=\bigl(\Delta_a^{(0)}\bigr)^{-1}-N\Pi_f,\\
&\Pi_b
=-\Delta_b^{-1}/N
=\Pi_\psi+\Pi_{\mathrm{CS}}
+N\Pi_{\mathrm{CS}}\!\cdot\!\Delta_a\!\cdot\!\Pi_{\mathrm{CS}}.
\end{aligned}
\right.
\label{eq:ha}
\end{equation}
The projected kernels use inverses on the gauge-transverse subspace.
Products denote tensor contractions.
The term $+\Pi_{\mathrm{CS}}$ in $\Pi_b$ comes from the level-$N$ Chern--Simons term of $b$.

Equation~\eqref{eq:ha} shows that the low-energy dynamics of \(b\) is inherited from two ingredients: the direct response of the Dirac sector, encoded in $\Pi_\psi$, and the indirect contribution mediated through the \(a\) gauge field, encoded in $\Pi_f$ and the Chern--Simons mixing. Solving these equations therefore yields the dressed kernel $\Pi_b$ entering the effective action for the critical gauge mode.

The internal fields are integrated out in a covariant gauge; the resulting $\Pi_b$ is transverse, so the gauge of the remaining field $b$ can be fixed independently. To separate the temporal and spatial components of \(b\), we use Coulomb gauge, $\nabla\!\cdot\!\boldsymbol b=0$. In this gauge the spatial component reduces to a single transverse scalar mode,
$\boldsymbol b(\boldsymbol q)
=(\hat{\boldsymbol z}\times\hat{\boldsymbol q})b(\boldsymbol q)$,
and the effective action becomes
\begin{equation}
\begin{aligned}
\mathcal S_b
=
-\frac{N}{2}\!\int_{\Omega_n,\boldsymbol q}
\Bigg(
\frac{q^2}{\Omega_n^2+q^2}
|b_0|^2\Pi_{b,L}
+|b|^2\Pi_{b,T} -\frac{q}{\sqrt{\Omega_n^2+q^2}}
(b_0^*b-b^*b_0)\Pi_{b,O}
\Bigg).
\end{aligned}
\end{equation}
Here $\Pi_{b,T}$, $\Pi_{b,L}$, and $\Pi_{b,O}$ are respectively the transverse, longitudinal, and parity-odd projected components of the polarization tensor. Their precise definitions and the evaluation of the corresponding finite-$\mu$ projected kernels are given in Appendix~\ref{app:kernels}.
The last term mixes $b_0$ and the transverse spatial mode $b$ through the parity-odd response $\Pi_{b,O}$, with the prefactor $q/\sqrt{\Omega_n^2+q^2}$ arising from the projection of the odd tensor structure onto the Coulomb-gauge basis.

After analytic continuation to real frequency we write $P$ for the retarded branch of the square root,
$P^2=q^2-(\Omega+i0^+)^2$, and the retarded propagator, with the overall factor
of $1/N$ removed, is
\begin{equation}
\mathcal G^R
=
\frac1{\Pi_{b,L}\Pi_{b,T}+\Pi_{b,O}^2}
\begin{pmatrix}
-\dfrac{P^2}{q^2}\Pi_{b,T}&
i\dfrac{P}{q}\Pi_{b,O}\\[2mm]
-i\dfrac{P}{q}\Pi_{b,O}&\Pi_{b,L}
\end{pmatrix},
\label{eq:Grets}
\end{equation}
with the projected kernels evaluated on the retarded branch. The overall factor of $1/N$ follows from $\Delta_b^{-1}=-N\Pi_b$ in Eq.~\eqref{eq:ha}; Eq.~\eqref{eq:Grets} is the standard analytic continuation of $\Delta_b$, including the Wick rotation of its temporal component. Its density and transverse diagonal entries define the positive-frequency spectral weights
\begin{align}
\label{eq:D0def}
D_0&=-\operatorname{Im}\mathcal G_{00}^R,\\
\label{eq:DTdef}
D_T&=-\operatorname{Im}\mathcal G_{TT}^R.
\end{align}
The leading collision kernel is obtained by contracting the matrix in Eq.~\eqref{eq:Grets} with the density and transverse-current fermion vertices. In the massless, particle--hole-symmetric approximation used here, the explicit mixed interference term in the on-shell weight vanishes, but both density and transverse diagonal channels remain, and the Chern--Simons mixing is kept inside the denominator $\Pi_{b,L}\Pi_{b,T}+\Pi_{b,O}^2$ of Eq.~\eqref{eq:Grets}. The resulting on-shell weights are constructed in Sec.~\ref{sec:qbe}. Under the opposite sign convention for the transverse basis vector, the off-diagonal entries of Eq.~\eqref{eq:Grets} and the transverse vertex change sign simultaneously, and all final scattering weights are unchanged.

The large-$\mu$, fixed-frequency limit of the projected kernels is derived in Appendix~\ref{app:kernels}. For $\Omega,q,T\ll\mu$ at fixed nonzero frequency, we write $\Psi_{L,T}\equiv\Pi_{\psi,L,T}$ for the retarded Dirac polarizations and $\bar\Psi_{L,T}(q)\equiv\Psi_{L,T}(0,q)$ for their static values. The two diagonal entries reduce to
\begin{equation}
\begin{aligned}
\mathcal G_{TT}^{(\infty)}
&=\frac{1}{\Psi_T}\left[1-\frac{P^2}{n_\psi\mathcal C}\right],\\
\mathcal G_{00}^{(\infty)}
&=-\frac{P^2}{q^2\Psi_L}\left[1-\frac{P^2}{n_\psi\mathcal C}\right],\\
D_{0,T}^{(\infty)}
&=-\operatorname{Im}\mathcal G_{00,TT}^{(\infty)},\\
\mathcal C&=P^2+4\pi^2\Psi_L\Psi_T,
\end{aligned}
\label{eq:Dinf}
\end{equation}
with $n_\psi=1$ at $\nu=-1/2$ and $n_\psi=2$ for the quarter-filled reference theory at $\nu=-3/4$. These are the regular spectra used in the collision integral; both diagonal entries are kept.

Figure~\ref{fig:spectra} shows these regular spectra for both fillings. Fix the spinon band parameters and take $q/T=O(1)$ and $\Omega/T\ne0$ fixed, followed by $T/\mu\to0$. In this fixed-frequency limit, neither diagonal spectrum of Eq.~\eqref{eq:Dinf} depends explicitly on the spinon parameters at leading order, and the density channel is not parametrically suppressed by $T/\mu$ at thermal momenta. This is not a statement that the two spectra are equal in all kinematic regions. The limit is not uniform near $\Omega=0$, where the spinon response must be retained before inverting the gauge kernel.

\subsection{Regular and quasistatic spectra}
The large-$\mu$ limit is not uniform near $\Omega=0$. Spinon Landau damping [Eq.~\eqref{eq:lindhard}] introduces the frequency scale $\Omega\sim T^2/\mu$, and the collision integral weights the spectrum by $n_{\mathrm{B}}(\Omega)\simeq T/\Omega$, so the relevant quantity is the Bose-weighted transverse spectrum $T D_T/\Omega$. As in the Mott-transition calculation of Witczak-Krempa \emph{et al.}~\cite{UniversalTransportQuantumwitczak-krempa2012}, keeping the spinon response in the projected kernels before inversion (Appendix~\ref{app:kernels}) yields, in addition to the smooth fixed-frequency part, a Lorentzian peak centered at $\Omega=0$,
\begin{equation}
\label{eq:Wsing}
\left.\frac{T}{\Omega}D_T(\Omega,q)\right|_{\mathrm{sing}}
=
\mathcal A(q)\,
\frac{\Omega_*(q)}{\Omega^2+\Omega_*(q)^2},
\end{equation}
with
\begin{align}
\label{eq:An}
\mathcal A(q)
&=\frac{Tq^2}
{n_\psi\,|\bar\Psi_T(q)|\,\bar{\mathcal C}(q)},\\
\label{eq:Gamman}
\Omega_*(q)
&=\frac{\pi v_{\mathrm{F}}q^3\,|\bar\Psi_T(q)|}
{n_\psi\,\mu\,\bar{\mathcal C}(q)},\\
\label{eq:c_q}
\zeta(q)
&=\frac{2\pi\bar\Psi_T(q)}{q},\\
\label{eq:Cdef}
\bar{\mathcal C}(q)
&=q^2+4\pi^2\bar\Psi_L(q)\bar\Psi_T(q).
\end{align}
Here $\bar\Psi_{L/T}(q)$ are the static Dirac polarizations, which are both negative so that $\bar{\mathcal C}(q)>0$, and $\bar{\mathcal C}(q)$ is the static value of $\mathcal C$ in Eq.~\eqref{eq:Dinf}. The density peak has the same width $\Omega_*(q)$ and carries weight $\zeta(q)^2\mathcal A(q)$, so that the singular parts satisfy
\begin{equation}
D_0=\zeta(q)^2D_T
\end{equation}
inside the window.

For $q\sim T$ the peak has height
$\mathcal A(q)/\Omega_*(q)=\mu T/(\pi v_{\mathrm{F}}q\,\bar\Psi_T^2)\sim\mu/T^2$
and width $\Omega_*\sim T^2/\mu$, while its weight is finite,
\begin{equation}
\label{eq:peak_weight}
\!\int_0^\infty\frac{\mathrm d\Omega}{\pi}\,
\left.\frac{T}{\Omega}D_T\right|_{\mathrm{sing}}
=\frac{\mathcal A(q)}{2}.
\end{equation}
As $\mu\to\infty$, the singular part tends to
$\pi\mathcal A(q)\delta(\Omega)$ on the full frequency axis.
The weight $\mathcal A(q)$ is independent of $\mu$ and
$v_{\mathrm{F}}$ and is determined by the Dirac polarizations alone;
the singular distribution appears in $TD_T/\Omega$, not in $D_T$
itself.

The regular and quasistatic (elastic) parts of the collision integral built from these spectra are constructed in Sec.~\ref{sec:qbe}.

\section{Quantum Boltzmann equation and critical transport at $\nu=-1/2$}
\label{sec:qbe}

We now compute the universal low-frequency transport in the quantum-critical regime. The Dirac-sector conductivity takes the scaling form~\cite{NonzerotemperatureTransportFractionalsachdev1998a}
\begin{equation}
\label{eq:sigscal}
\sigma_\psi(\omega,T)
=
{2\pi\frac{e^2}{h}}
\Sigma\!\left(
\frac{\hbar\omega}{k_BT}
\right).
\end{equation}
At leading order the scaling function consists of an intraband part, obtained from the QBE and denoted $\Sigma_{\mathrm{QBE}}(\omega/T)$, and the collisionless interband part of Eq.~\eqref{eq:re_coh}. The latter vanishes as $\omega\to0$, so $\Sigma(0)=\Sigma_{\mathrm{QBE}}(0)$. With $\hbar=1$, the QBE result is expressed in units of $e^2/\hbar$; the physical conductance in units of $e^2/h$ is $2\pi\Sigma_{\mathrm{QBE}}$, since $e^2/h=e^2/(2\pi\hbar)$.
Its collisionless limit $\Sigma(\infty)$ is given by Eq.~\eqref{eq:inf_limit}, whereas the DC value $\Sigma(0)$ is set by collisions of thermally excited carriers with gauge fluctuations, both inelastic and quasistatic. These collisions are encoded in the Coulomb-gauge density/transverse-current propagator of Eq.~\eqref{eq:Grets}, contracted with the fermion density and transverse-current vertices (Sec.~\ref{sec:gauge} and Appendix~\ref{app:kernels}). To compute $\Sigma$, we couple an auxiliary probe field to one replica $\psi_1$ of the two-cone Dirac sector; the remaining $N-1$ replicas stay in equilibrium and act as a bath. Since the gauge propagator is $O(1/N)$, the transport rate scales as
\begin{equation}
\label{eq:sct_rate}
\frac{1}{\tau_{\mathrm{tr}}}
\sim
\frac{T}{N},
\end{equation}
up to a universal coefficient. These collisions broaden the $N=\infty$ Drude singularity into a peak of width $\sim T/N$ and render $\sigma_\psi(0)$ finite at the leading nontrivial order.

\subsection{Kinetic setup}

We formulate the quantum Boltzmann equation appropriate for the low-frequency transport problem. Our use of a quantum Boltzmann equation is also motivated by earlier kinetic treatments of composite fermions coupled to gauge fields, where gauge-boson scattering was shown to control low-frequency transport~\cite{PhysRevB.52.17275}. To probe \(\sigma_\psi\), we apply a spatially uniform auxiliary electric field \(\boldsymbol E(t)\) oscillating at frequency \(\omega\), keeping $\omega/T$ fixed before taking $\omega\to0$. The probe drives $\psi_1$, whose relaxation is controlled by scattering from the gauge fluctuations of $b$.

To set up the kinetic description, it is convenient to expand \(\psi_1\) in the single-particle eigenbasis,
\begin{equation}
\psi_1(\boldsymbol x,t)
=
\!\int_{\boldsymbol k}
\frac{1}{\sqrt{2}}\left[
\gamma_+(\boldsymbol k,t)
\begin{pmatrix}
1\\
e^{i\phi_{\boldsymbol k}}
\end{pmatrix}
+
\gamma_-(\boldsymbol k,t)
\begin{pmatrix}
1\\
-e^{i\phi_{\boldsymbol k}}
\end{pmatrix}
\right]
e^{i\boldsymbol k\!\cdot\!\boldsymbol x},
\end{equation}
where \(\phi_{\boldsymbol k}=\arg(k_x+ik_y)\). The distribution $f_-$ below is the negative-band electron occupation. In the kinetic description below, the physical cone index of \(f_s\) and \(\gamma_s\) is suppressed, and all cones have the same distribution~\cite{NonzerotemperatureTransportFractionalsachdev1998a}.

In this basis, the kinetic current is \(J_i=\psi_1^\dagger\sigma^i\psi_1=-i\bar\psi_1\gamma^i\psi_1\), and it naturally separates into an intraband contribution and an interband coherence contribution, \(\boldsymbol J=\boldsymbol J_{\mathrm I}+\boldsymbol J_{\mathrm{II}}\). The intraband part is
\begin{equation}
\begin{aligned}
\boldsymbol J_{\mathrm I}(t)
&=
\frac{2}{V}\sum_{s=\pm}
\!\int_{\boldsymbol k}
\frac{s\boldsymbol k}{\varepsilon_{\boldsymbol k}}
\langle
\gamma_s^\dagger(\boldsymbol k,t)
\gamma_s(\boldsymbol k,t)
\rangle,
\label{eq:j_from_f}
\end{aligned}
\end{equation}
while the interband term is
\begin{equation}
\boldsymbol J_{\mathrm{II}}
=
\frac{2i}{V}\!\int_{\boldsymbol k}
\frac{\hat{\boldsymbol z}\times\boldsymbol k}{|\boldsymbol k|}
\left\langle
\gamma_+^\dagger(\boldsymbol k)\gamma_-(\boldsymbol k)
-
\gamma_-^\dagger(\boldsymbol k)\gamma_+(\boldsymbol k)
\right\rangle.
\end{equation}
The term \(\boldsymbol J_{\mathrm I}\) is the conventional transport current carried by Dirac quasiparticles in the positive- and negative-energy bands, with opposite signs reflecting their opposite band velocities. By contrast, $\boldsymbol J_{\mathrm{II}}$ represents an interband coherence contribution associated with particle--hole pair creation. We are interested in the low-frequency transport regime, where the current is dominated by the intraband dynamics of thermally excited carriers, so we work in the approximation $\boldsymbol J_{\mathrm{II}}=0$.

We now define the distribution functions
\begin{equation}
\label{eq:fs}
f_s(\boldsymbol k,t)
=
\frac{1}{V}
\langle
\gamma_s^\dagger(\boldsymbol k,t)
\gamma_s(\boldsymbol k,t)
\rangle,
\end{equation}
where \(V\) denotes the system volume, which may equivalently be viewed as the \(\delta_{\boldsymbol k}(0)\) factor arising in momentum space. In thermal equilibrium and in the absence of external perturbations, the distribution functions reduce to the Fermi--Dirac forms
\begin{equation}
f_\pm^{(0)}(\boldsymbol k)
=
n_{\mathrm F}(\pm\varepsilon_k)
=
\frac{1}{e^{\pm\varepsilon_k/T}+1},
\qquad
\varepsilon_k=|\boldsymbol k|.
\end{equation}

\subsection{Collision integral}

The collision integral follows from the Coulomb-gauge propagator of Eq.~\eqref{eq:Grets}, contracted with the density and transverse-current vertices: the interaction $H_{\mathrm{int}}=b_0\rho-b_Tj_T$ couples $b_0$ to the fermion density $\rho$ and $b_T$ to the transverse current $j_T$, and for the single-particle states $u_s(\boldsymbol k)=(1,\,se^{i\phi_{\boldsymbol k}})^{\mathsf T}/\sqrt2$ with the transverse polarization vector $\hat{\boldsymbol e}_T=\hat{\boldsymbol z}\times\hat{\boldsymbol q}$, the overlap amplitudes are $r_{ss'}=u_{s'}^\dagger(\boldsymbol k')u_s(\boldsymbol k)$ and $t_{ss'}=u_{s'}^\dagger(\boldsymbol k')(\boldsymbol\sigma\!\cdot\!\hat{\boldsymbol e}_T)u_s(\boldsymbol k)$. The mixed term between the two channels cancels in the massless, particle--hole-symmetric approximation, so that the transition weight is the sum of the density and transverse-current channels,
\begin{equation}
\label{eq:Wpm}
W_\pm(\Omega;k,k',\theta)
=
\frac{1\pm\cos\theta}{2}D_0(\Omega,q)
+
\frac{(k\pm k')^2(1\mp\cos\theta)}{2q^2}D_T(\Omega,q),
\end{equation}
with $q^2=k^2+k'^2-2kk'\cos\theta$ and, on the energy shell, $\Omega=|k\mp k'|$. Here $D_0$ and $D_T$ are the density and transverse spectra of the retarded propagator of Eq.~\eqref{eq:Grets}, and $W_{ss'}=W_+$ for $s'=s$ while $W_{ss'}=W_-$ for $s'=-s$. The cancellation of the mixed term is a property of the massless, particle--hole-symmetric approximation used here, not a statement that interference is absent in massive or higher-order theories. The Chern--Simons mixing is not removed by this cancellation: it remains encoded in the denominators of $D_0$ and $D_T$ through Eq.~\eqref{eq:Grets}. For a physical spectrum the on-shell weights $W_{ss'}$ are nonnegative.

In the regular part of the collision integral below, $D_0$ and $D_T$ are replaced by their fixed-frequency limits $D_{0,T}^{(\infty)}$ of Eq.~\eqref{eq:Dinf}. Applying Fermi's golden rule, the collision term can be written as

\begin{equation}
\begin{aligned}
(\partial_t+\boldsymbol E\!\cdot\!\partial_{\boldsymbol k})f_s(\boldsymbol k)
={}&-\frac1N\sum_{s'=\pm}
\!\int_0^\infty\frac{\mathrm d\Omega}{\pi}
\!\int\!\frac{\mathrm d^2q}{(2\pi)^2}
W_{ss'}(\Omega;\boldsymbol k,\boldsymbol k+\boldsymbol q)
(2\pi)
\\
&\times\Big\{
\delta(s\varepsilon_k-s'\varepsilon_{k+q}-\Omega)
\Big[
f_s(\boldsymbol k)
(1-f_{s'}(\boldsymbol k+\boldsymbol q))
(1+n_{\mathrm{B}}(\Omega))
\\
&\hspace{46mm}
-f_{s'}(\boldsymbol k+\boldsymbol q)
(1-f_s(\boldsymbol k))
n_{\mathrm{B}}(\Omega)
\Big]
\\
&\quad+
\delta(s\varepsilon_k-s'\varepsilon_{k+q}+\Omega)
\Big[
f_s(\boldsymbol k)
(1-f_{s'}(\boldsymbol k+\boldsymbol q))
n_{\mathrm{B}}(\Omega)
\\
&\hspace{46mm}
-f_{s'}(\boldsymbol k+\boldsymbol q)
(1-f_s(\boldsymbol k))
(1+n_{\mathrm{B}}(\Omega))
\Big]
\Big\}.
\end{aligned}
\label{eq:qbe}
\end{equation}

Here \(n_{\mathrm{B}}(\Omega)\) is the Bose distribution, the internal energy transfer \(\Omega\) is distinct from the external probe frequency \(\omega\), and the explicit factor \(1/N\) is kept outside the collision integral.

The frequency integral in Eq.~\eqref{eq:qbe} receives two contributions (Sec.~\ref{sec:gauge}). Outside the window $\Omega\sim T^2/\mu$, the diagonal spectra are given by their fixed-frequency limits $D_{0,T}^{(\infty)}$ of Eq.~\eqref{eq:Dinf}. Inside the window, $n_{\mathrm{B}}(\Omega)\simeq T/\Omega$, the emission and absorption terms add, and $T D_T/\Omega$ develops the quasistatic peak of Sec.~\ref{sec:gauge}: a Lorentzian with weight $\mathcal A(q)$ and width $\Omega_*(q)$ [Eqs.~\eqref{eq:An} and~\eqref{eq:Gamman}], whose density component carries an additional factor $\zeta(q)^2$ [Eq.~\eqref{eq:c_q}]. On the elastic shell $k'=k$, the contraction with the on-shell vertices and the integrated peak weight gives $\mathcal A(q)[1+\zeta(q)^2\cos^2(\theta/2)]$. Only intraband elastic processes survive in the quasistatic limit, and the Pauli factors cancel, $f_s(1-f_{s'})-f_{s'}(1-f_s)=f_s-f_{s'}$. This gives the additional collision term
\begin{equation}
\label{eq:Iel}
\begin{aligned}
I_{\mathrm{el}}[f_s](\boldsymbol k)
=-\frac{2\pi}{N}
\!\int\!\frac{\mathrm d^2k'}{(2\pi)^2}
\,\delta(k-k')
\mathcal A(q)\left[1+\zeta(q)^2\cos^2\frac\theta2\right]
[f_s(\boldsymbol k)-f_s(\boldsymbol k')],
\end{aligned}
\end{equation}
with $q=|\boldsymbol k'-\boldsymbol k|$, which is added to the right-hand side of Eq.~\eqref{eq:qbe}.

The gauge spectrum is evaluated in equilibrium: nonequilibrium corrections from the single driven replica are subleading in $1/N$, and the gauge fluctuations remain described by the equilibrium propagator of Eq.~\eqref{eq:Grets}.

\subsection{Linearization and scaling form}

To compute the conductivity of the critical Dirac sector, we solve the quantum Boltzmann equation in the presence of a weak auxiliary electric field \(\boldsymbol E\). Near equilibrium, the distribution function \(f_s(\boldsymbol k,\omega)\) for band index \(s=\pm\) can be expanded to linear order in \(\boldsymbol E\) as
\begin{equation}
\label{eq:fslinear}
f_s(\boldsymbol k,\omega)
=
n_{\mathrm F}(s\varepsilon_k)\,2\pi\delta(\omega)
+
s\boldsymbol E\!\cdot\!\boldsymbol k\,
\varphi(k,\omega),
\end{equation}
where we have Fourier transformed from time to frequency space. The first term is the equilibrium Fermi--Dirac distribution, while the second term describes the anisotropic deformation induced by the electric field; the scalar function \(\varphi(k,\omega)\) parametrizes the linear response of the electron distribution. At Dirac neutrality, within the leading particle--hole-symmetric kinetic approximation, the field drives a counterflow with vanishing total Dirac momentum but nonzero electrical current. Collisions can therefore relax this current without requiring momentum relaxation~\cite{NonzerotemperatureTransportQuantumdamle1997,NonzerotemperatureTransportFractionalsachdev1998a}.

Substituting Eqs.~\eqref{eq:fslinear} and~\eqref{eq:fs} into the current, Eq.~\eqref{eq:j_from_f}, gives the conductivity of the Dirac sector. After linearization it takes the form given below.
The collision integral can then be linearized around equilibrium; the details are given in Appendix~\ref{app:collision}. The resulting kinetic equation for \(\varphi\) takes the form
\begin{equation}
\label{eq:qbe1}
\begin{aligned}
-i\omega\varphi(k,\omega)
+\frac{g(k/T)}{T^2}
={}&
\frac{T}{N}\Big\{
-\big[F_b(k/T)+F_{\mathrm{el}}(k/T)\big]\varphi(k,\omega)\\
&+\!\int\!\frac{\mathrm dk'}{T}K_b(k'/T,k/T)\varphi(k',\omega)
\Big\}
\end{aligned}
\end{equation}
Here $F_b$ and $K_b$ are built from the on-shell weights $W_\pm$ (Appendix~\ref{app:collision}), and
\begin{equation}
\label{eq:Fel}
F_{\mathrm{el}}(k)
=
k\!\int_0^{2\pi}\frac{\mathrm d\theta}{2\pi}
(1-\cos\theta)\,
\mathcal A(q)
\left[
1+\zeta(q)^2\cos^2\frac\theta2
\right],
\qquad
q=2k\left|\sin\frac\theta2\right|,
\end{equation}
with all momenta measured in units of $T$, and with $\mathcal A(q)$ and $\zeta(q)$ the quasistatic weight and the density-to-transverse ratio of Sec.~\ref{sec:gauge} [Eqs.~\eqref{eq:An} and~\eqref{eq:c_q}]. Equation~\eqref{eq:Fel} combines elastic out-scattering and in-scattering through the factor $1-\cos\theta$.

The source term on the left-hand side is
\begin{equation}
g(k)
=
\frac{1}{k}
\frac{\mathrm d}{\mathrm dk}
\frac{1}{e^{k}+1}
=
-\frac{e^{k}}
{k(e^{k}+1)^2},
\end{equation}
which arises from the linearized drift term \(\boldsymbol E\!\cdot\!\partial_{\boldsymbol k}f_s\) evaluated on the equilibrium distribution; here $k$ is measured in units of $T$.

The two terms on the right-hand side are the out-scattering and in-scattering parts of the same linearized collision operator. The diagonal term proportional to \(F_b\) collects contributions multiplying the distribution distortion at \(k\), whereas \(K_b(k',k)\) couples that distortion to those at other momenta; both are built from the weights $W_\pm$ and include energy-exchanging processes. The individual diagonal contribution $F_b$ can contain forward singularities; the physical relaxation scale is set by the collision operator as a whole, after the out-scattering and in-scattering terms are combined with a common infrared regulator, with the parametric form of Eq.~\eqref{eq:sct_rate}. We solve the linearized equation including $K_b$. Because the fixed-frequency spectra $D_{0,T}^{(\infty)}$ and $\mathcal A(q)$ depend only on the Dirac polarization, $F_b$, $K_b$, and $F_{\mathrm{el}}$ are universal.

To expose the universal scaling structure in the quantum critical regime, we measure fermion momenta in units of $T$ and introduce
\begin{equation}
\tilde\omega=\frac{\omega N}{T},
\qquad
\chi(k,\tilde\omega)
=
\frac{T^3}{N}\varphi(Tk,\omega),
\end{equation}
where the momentum argument of $\chi$ is in units of $T$. In terms of these variables, Eq.~\eqref{eq:qbe1} becomes
\begin{equation}
\label{eq:qbe_dimensionless}
-i\tilde\omega\chi(k,\tilde\omega)
+g(k)
=
-\big[F_b(k)+F_{\mathrm{el}}(k)\big]\chi(k,\tilde\omega)
+\!\int\!\mathrm dk'\,K_b(k',k)\chi(k',\tilde\omega).
\end{equation}
This dimensionless form makes explicit that, at fixed \(N\), the conductivity is controlled by a universal scaling function of \(\omega/T\).

Taking the auxiliary electric field along the \(x\)-direction, the conductivity can be written as~\cite{UniversalTransportQuantumwitczak-krempa2012}
\begin{equation}
\label{eq:sig_psi}
\sigma_\psi(\omega)
=
\frac{N}{\pi}
\!\int_0^{\Lambda/T}
\mathrm dk\,
k^2\chi(k,\tilde\omega).
\end{equation}
The variable $\tilde\omega=N\omega/T$ makes the $\omega/T$ scaling at fixed $N$ explicit.

\subsection{Results for \texorpdfstring{$\nu=-1/2$}{nu=-1/2}}

\begin{figure}[t]
    \centering
    \includegraphics[width=0.92\textwidth]{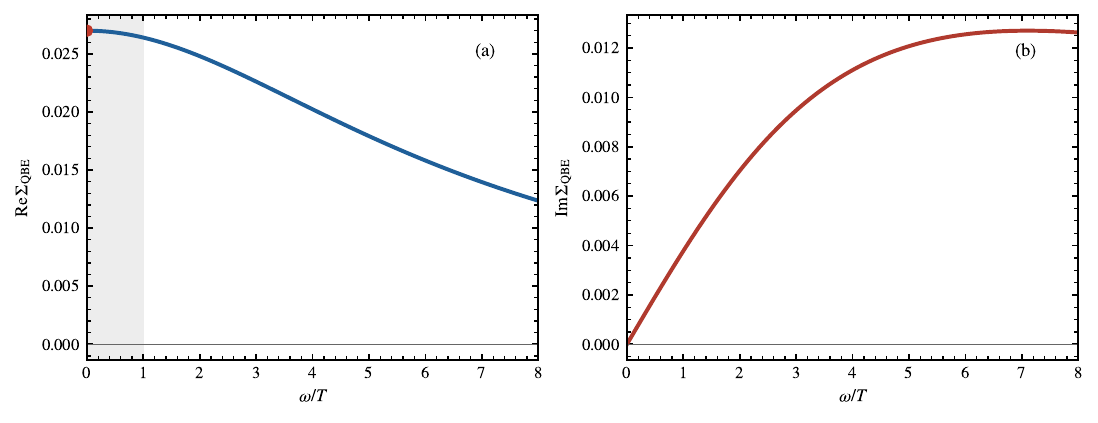}
    \caption{Dirac-sector scaling function $\Sigma_{\mathrm{QBE}}(\omega/T)$ from the intraband QBE at $N=1$ and $\nu=-1/2$, computed with the complete density/transverse collision kernel. (a) Real part; the filled point marks the DC value and the shaded region is $\omega/T<1$. (b) Imaginary part.}
    \label{fig:sig12}
\end{figure}

Setting $N=1$, we solve Eq.~\eqref{eq:qbe_dimensionless} numerically for the intraband part of $\Sigma(\omega/T)$, shown in Fig.~\ref{fig:sig12}.

Its DC limit gives
\begin{equation}
\label{eq:sigma_psi_DC}
\sigma_\psi(0)
=
0.1696\,\frac{e^2}{h},
\end{equation}
which enters the physical resistivity through Eq.~\eqref{eq:rho_explicit} (Sec.~\ref{sec:experiment}).

\section{Critical transport at $\nu=-3/4$}
\label{sec:34}

We now extend the transport analysis to the CFL--FL transition at filling \(\nu=-3/4\).
The critical theory is more involved than at $\nu=-1/2$, but the transport calculation follows the same logic: we first derive the Ioffe--Larkin composition rule for the reference resistivity, then identify the effective low-energy gauge propagator that scatters the critical Dirac fermions, and finally solve the corresponding quantum Boltzmann equation to obtain the universal quantum-critical conductivity.

For $\nu=-3/4$, we use the quarter-filled CFL--FL reference theory whose particle--hole conjugate describes the experimental filling.
Its charged boson undergoes a Laughlin--superfluid transition at effective filling $\nu_\Phi=1/4$, coupled to a neutral Fermi surface through emergent gauge fields.
The Lagrangian and Ioffe--Larkin rule below refer to this quarter-filled reference response; the physical conductivity is reconstructed from it by particle--hole conjugation, and the reconstruction includes the Hall contribution of the filled $C=1$ band. This filled band is part of the same band, not a separate parallel band.
A convenient parton construction for the reference theory is~\cite{song_2024phase}
\begin{equation}
\Phi=f_1f_2f_3f_4,
\end{equation}
which introduces additional gauge constraints.
In the mean-field construction of Ref.~\cite{song_2024phase}, tuning the bandwidth drives a Chern-number-changing transition in the \(f_3,f_4\) sector.
At low energies this produces two two-cone Dirac sectors ($n_\psi=2$), coupled to emergent gauge fields \(b_i\). By the exchange symmetry of the two sectors it suffices to consider $b\equiv b_1$. The critical theory is therefore described by
\begin{equation}
\mathcal L_{\mathrm{crit}}
=
\mathcal L_\psi+\mathcal L_f+\mathcal L_{\mathrm{CS}},
\end{equation}
with
\begin{equation*}
\begin{aligned}
\mathcal L_\psi
&=
\sum_{i=1}^{2}
\left[
-i\overline{\psi_i}
(\slashed{\partial}\tau_0-i\slashed{b}_i\tau_0)\psi_i
+M\overline{\psi_i}\psi_i
\right],\\
\mathcal L_f
&=
f^\dagger
\left[
i\partial_t-a_0+\mu
+\frac{1}{2m}(\nabla+i\boldsymbol a)^2
\right]f,\\
\mathcal L_{\mathrm{CS}}
&=
\frac{1}{2\pi}(A+a+b_1+b_2)\,\mathrm d\alpha
-\frac{2}{4\pi}\alpha\,\mathrm d\alpha
-\sum_{i=1,2}\frac{1}{4\pi}b_i\,\mathrm db_i.
\end{aligned}
\end{equation*}
Each two-cone sector carries the same two-component \(\tau\)-space structure. Lattice translations interchange the two sectors, reflecting the nontrivial symmetry implementation in the underlying moir\'e system.

As at $\nu=-1/2$, the contact terms cancel the explicit $-\frac{1}{4\pi}b_i\,\mathrm db_i$ terms. Integrating out the matter fields gives the Gaussian effective action
\begin{multline}
\mathcal S_E[b_i,a,\alpha;A]
=
-\frac12\!\int\!
\Bigg[
\sum_{i=1,2}
b_{i\mu}^*
\Pi_{\psi_i}^{\mu\nu}b_{i\nu}
+a_\mu^*\Pi_f^{\mu\nu}a_\nu\\
-2\alpha_\mu^*\Pi_{\mathrm{CS}}^{\mu\nu}\alpha_\nu
+2(A_\mu+a_\mu+b_{1\mu}+b_{2\mu})^*
\Pi_{\mathrm{CS}}^{\mu\nu}\alpha_\nu
\Bigg].
\end{multline}
The electromagnetic response of the quarter-filled reference theory follows from the linear saddle-point equations for the internal gauge fields,
\begin{equation}
\left\{
\begin{aligned}
&-\frac{\delta\mathcal S_E}{\delta b_{i\mu}^*}
=
\Pi_{\psi_i}^{\mu\nu}b_{i\nu}
+\Pi_{\mathrm{CS}}^{\mu\nu}\alpha_\nu=0,\\
&-\frac{\delta\mathcal S_E}{\delta a_\mu^*}
=
\Pi_f^{\mu\nu}a_\nu
+\Pi_{\mathrm{CS}}^{\mu\nu}\alpha_\nu=0,\\
&-\frac{\delta\mathcal S_E}{\delta\alpha_\mu^*}
=
-2\Pi_{\mathrm{CS}}^{\mu\nu}\alpha_\nu
+\Pi_{\mathrm{CS}}^{\mu\nu}
(A_\nu+a_\nu+b_{1\nu}+b_{2\nu})=0.
\end{aligned}
\right.
\end{equation}
Solving these equations yields the Ioffe--Larkin rule for the quarter-filled reference polarization tensor,
\begin{equation}
\label{eq:ilre}
(\Pi^{\mathrm{ref}})^{-1}
=
2\Pi_{\mathrm{CS}}^{-1}
+\Pi_f^{-1}
+\Pi_{\psi_1}^{-1}
+\Pi_{\psi_2}^{-1}.
\end{equation}
In the isotropic transport limit, the charged-sector tensor of the reference theory is
\begin{equation}
\label{eq:resis34}
(\rho_\Phi)_{ij}
=
\frac{2}{\sigma_\psi}\delta_{ij}
-4\pi\epsilon_{ij}.
\end{equation}
Here $\sigma_\psi$ is the conductivity of one two-cone Dirac
sector, common to both sectors, and its multiplicity $n_\psi=2$
produces the overall factor relative to Eq.~\eqref{eq:rho_explicit}.
The series combination $\rho_f\mathbb I+\rho_\Phi$ is the
quarter-filled reference resistivity, not the measured
$\nu=-3/4$ tensor, which follows from Eq.~\eqref{eq:rho_phys_iso}.

We next generalize the large-\(N\) construction. As before, we introduce \(N\) copies of each gapless matter field and scale the Chern--Simons levels with \(N\), so that the gauge-field dynamics remains nontrivial in the \(1/N\) expansion. The corresponding Euclidean Lagrangian is
\begin{equation}
\mathcal L_N
=
\sum_{l=1}^N
\left(
\mathcal L_{\psi,l}+\mathcal L_{f,l}
\right)
-\frac{iN}{2\pi}(a+b_1+b_2)\mathrm d\alpha
+\frac{2iN}{4\pi}\alpha\mathrm d\alpha.
\end{equation}
The number of Dirac sectors is enlarged from \(2\) to \(2N\), with the \(\tau\)-space structure of each sector retained. We use the same Euclidean gamma-matrix convention as in Sec.~\ref{sec:largeN}.

At leading order in \(1/N\), the matter fields contribute through the same one-loop polarization bubbles \(\Pi_\psi\) and \(\Pi_f\) as at $\nu=-1/2$; the difference lies entirely in how these bubbles are reorganized by the Chern--Simons mixing.
The inverse of the bare $\alpha$ propagator in the covariant gauge of Sec.~\ref{sec:largeN} is
\begin{equation*}
\big[(\Delta_\alpha^{(0)})^{-1}\big]_{\mu\nu}
=2N(\Pi_{\mathrm{CS}})_{\mu\nu}+\frac{2N}{\xi}P_\mu P_\nu.
\end{equation*}
Following the same procedure as in Sec.~\ref{sec:gauge}, we integrate out the auxiliary gauge fields and obtain the effective kernel for the gauge mode \(b\) that scatters the critical Dirac fermions:
\begin{equation}
\left\{
\begin{aligned}
&\Delta_{b_2}^{-1}
=-N\Pi_\psi,
\qquad
\Delta_a^{-1}=-N\Pi_f,\\
&\Delta_\alpha^{-1}
=
\bigl(\Delta_\alpha^{(0)}\bigr)^{-1}
-N^2\Pi_{\mathrm{CS}}\!\cdot\!(\Delta_{b_2}+\Delta_a)\!\cdot\!\Pi_{\mathrm{CS}},\\
&N\Pi_b
=-\Delta_{b_1}^{-1}
=
N\Pi_\psi
+N^2\Pi_{\mathrm{CS}}\!\cdot\!\Delta_\alpha\!\cdot\!\Pi_{\mathrm{CS}}.
\end{aligned}
\right.
\label{eq:b34}
\end{equation}
Equation~\eqref{eq:b34} encodes the successive elimination of $a$, $b_2$, and $\alpha$.

Compared with the \(\nu=-1/2\) case, the effective propagator has a different functional form because of the modified gauge-field and Chern--Simons mixing structure, but it retains the form of Eq.~\eqref{eq:Grets}, with the projected kernels obtained from Eq.~\eqref{eq:b34} and the \(\nu=-3/4\) kernels of Appendix~\ref{app:kernels}. The regular spectra $D_T$ and $D_0$ are those of Eq.~\eqref{eq:Dinf} evaluated at $n_\psi=2$, the quasistatic peak is that of Sec.~\ref{sec:gauge} with the same $n_\psi$, and the linearized quantum Boltzmann equation of Sec.~\ref{sec:qbe} applies directly. The multiplicity $n_\psi=2$ enters only through the projected kernels, and hence through Eqs.~\eqref{eq:Dinf}, \eqref{eq:An}, and~\eqref{eq:Gamman}.

\begin{figure}[t]
    \centering
    \includegraphics[width=0.92\textwidth]{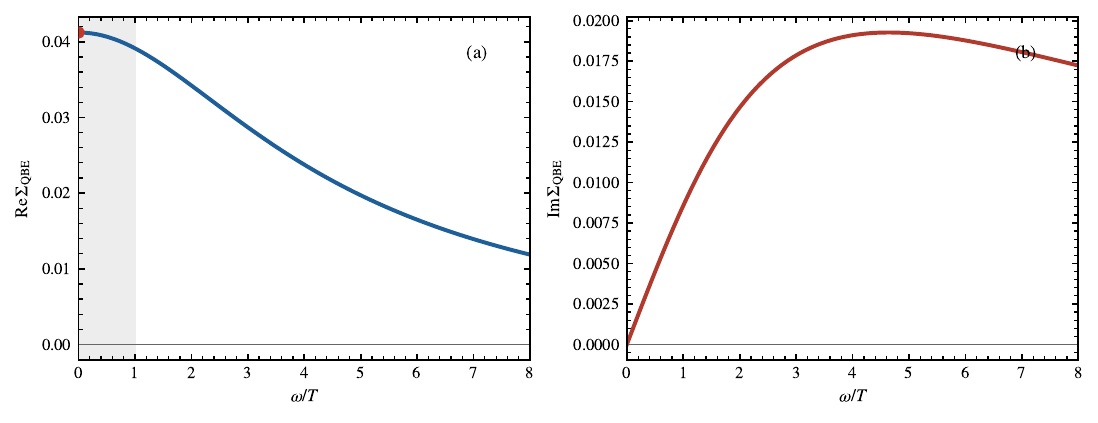}
    \caption{Dirac-sector scaling function $\Sigma_{\mathrm{QBE}}(\omega/T)$ from the intraband QBE at $N=1$ and $\nu=-3/4$, computed with the complete density/transverse collision kernel. (a) Real part; the filled point marks the DC value and the shaded region is $\omega/T<1$. (b) Imaginary part.}
    \label{fig:sig34}
\end{figure}

Solving this linearized equation yields the universal scaling function \(\Sigma(\omega/T)\) for this transition; its real part is shown in Fig.~\ref{fig:sig34}.

The DC limit gives
\begin{equation}
\label{eq:sigma_psi_DC34}
\sigma_\psi(0)
=
0.2591\,\frac{e^2}{h},
\end{equation}
which determines the reference charged-sector tensor through
Eq.~\eqref{eq:resis34} and the measured DC response through
Eq.~\eqref{eq:rho_phys_iso}.

\section{Discussion}
\label{sec:discussion}

\subsection{High-frequency conductivity}

The DC conductivities calculated in Secs.~\ref{sec:qbe} and \ref{sec:34} correspond to the hydrodynamic limit, $\hbar\omega\ll k_BT$, where scattering of the critical excitations off gauge fluctuations dominates. In this regime, the system behaves as an incoherent thermal fluid. However, for $\hbar\omega\gg k_BT$, the system enters a collisionless regime where transport is governed by the coherent creation of particle--hole pairs. In this optical limit, the longitudinal conductivity approaches the value of Eq.~\eqref{eq:inf_limit}, common to both cases~\cite{NonzerotemperatureTransportQuantumdamle1997}.

Between these limits, the Dirac conductivity crosses over
between its hydrodynamic and collisionless responses.
The intraband QBE results in Figs.~\ref{fig:sig12}
and~\ref{fig:sig34} describe the low-frequency contribution,
not the optical crossover.
The Ioffe--Larkin composition, Eq.~\eqref{eq:intro_IL}, converts the
frequency-dependent $\sigma_\psi$ into the charged-sector tensor
$\rho_\Phi(\omega)$.
At $\nu=-1/2$, this adds in series to
$\rho_f(\omega)\mathbb I$.
At $\nu=-3/4$, it is the charged-sector tensor of the
quarter-filled reference theory; the measured DC response
is reconstructed using Eq.~\eqref{eq:rho_phys_iso}.

\subsection{Zero-temperature decoupling versus finite-temperature transport}\label{sec:RG_decoupling}

Zero-temperature stability of the charged critical sector
does not by itself determine finite-temperature DC transport.
This distinction was emphasized for the Mott transition in
Refs.~\cite{TheoryContinuousMottsenthil2008a,
UniversalTransportQuantumwitczak-krempa2012}, and
Ref.~\cite{song_2024phase} anticipated an additional
finite-temperature scattering contribution from quasistatic
gauge fluctuations for the CFL--FL transition.
Retaining the spinon response in the projected $b$-field kernel
before inversion, together with the Dirac Chern--Simons contact
term, yields a zero-frequency peak in the Bose-weighted matrix
spectrum of the gauge mode that scatters the critical Dirac
fermions (Sec.~\ref{sec:gauge} and Appendix~\ref{app:kernels}).
Chern--Simons mixing changes the peak weight through
$\bar{\mathcal C}(q)$ and $n_\psi$ but does not remove the peak.
The peak therefore provides a quasistatic scattering channel at
finite temperature, and zero-temperature decoupling of the spinon
Fermi surface from the critical sector does not imply its
decoupling from finite-temperature DC transport.

\subsection{Thermal conductivity}

Thermal transport provides an additional diagnostic of the critical regime. Heat currents are not constrained by the internal gauge constraint, so, neglecting intersector drag and thermoelectric corrections, the sector contributions add in parallel~\cite{PhysRevB.46.5621},
\begin{equation}
\kappa\simeq\kappa_\psi+\kappa_f,
\end{equation}
in contrast to the series addition of electrical resistivities.

A decoupled, translation-invariant Dirac sector would be a CFT with a conserved energy current and hence a divergent DC thermal conductivity, cut off only by disorder or umklapp scattering. Here the Dirac heat current can relax by exchanging energy and momentum with the other sectors through the gauge fluctuations. If this process controls the relaxation and introduces no additional infrared scale, quantum-critical scaling gives
\begin{equation}
\kappa_\psi=\frac{k_B^2}{\hbar}K_\psi T,
\end{equation}
with a dimensionless coefficient $K_\psi$. The neutral Fermi surface carries heat but no charge; with elastic momentum relaxation it contributes the metallic form
\begin{equation}
\kappa_f=c_fT,
\end{equation}
where $c_f$ is set by the momentum-relaxation rate and by interaction corrections in the $f$--gauge sector~\cite{PhysRevB.76.235124}.

Because the sectors combine in series for charge but in parallel for heat, the Lorenz ratio
\begin{equation}
L^{\mathrm{phys}}
\equiv
\frac{\kappa}{T\sigma^{\mathrm{phys}}_{xx}}
\simeq
\frac{\kappa_\psi+\kappa_f}{T\sigma^{\mathrm{phys}}_{xx}},
\end{equation}
with $\sigma^{\mathrm{phys}}_{xx}$ from Sec.~\ref{sec:IL}, can deviate from the Wiedemann--Franz value. Determining $K_\psi$, $c_f$, and $L^{\mathrm{phys}}$ requires a consistent treatment of the coupled conserved modes and of momentum relaxation, which we leave to future work.

\subsection{Experimental implications}
\label{sec:experiment}

Our results apply to TMD homobilayers and graphene moir\'e
systems where the displacement field tunes the effective
bandwidth~\cite{lu2023fractional,park2023observation}.
We denote by $\rho$ the measured resistivity tensor and by
$\sigma^{\mathrm{phys}}=\rho^{-1}$ the physical conductivity.

The universal quantity is the clean-limit QCP--FL jump
$\Delta\rho$ defined in Eq.~\eqref{eq:r_definition}.
With the dimensionless Dirac conductance
\begin{equation}
\label{eq:gpsi_experiment}
g_\psi=\frac{h}{e^2}\sigma_\psi(0),
\end{equation}
the critical charged-sector tensor is
\begin{equation}
\label{eq:jump}
\rho_\Phi
=n_\psi\frac{h}{e^2}
\left[
\frac{1}{g_\psi}\mathbb I-\epsilon
\right],
\end{equation}
with $n_\psi=1$ $(2)$ at $\nu=-1/2$ $(-3/4)$.
At $\nu=-1/2$, $\rho=\rho_f\mathbb I+\rho_\Phi$ and the clean FL
has $\rho=0$; the resulting tensor reproduces Eq.~\eqref{eq:intro_jump} for $\nu=-1/2$.
At this order the Hall jump is fixed,
$\Delta\rho_{xy}=-h/e^2$, half of the CFL value $-2h/e^2$ in our sign convention, and only
$\Delta\rho_{xx}$ depends on $g_\psi$.
A Hall component of the dynamical Dirac response, absent at
leading order, would modify both components.

At $\nu=-3/4$, $\rho_f\mathbb I+\rho_\Phi$ is the resistivity of
the quarter-filled reference theory. Particle--hole conjugation
transposes its conductivity and adds the filled-band Hall
conductivity:
\begin{equation}
\label{eq:rho_phys_iso}
\rho=\left[
\sigma_{\mathrm{CS}}
+\left\{(\rho_f\mathbb I+\rho_\Phi)^{-1}\right\}^{\mathsf T}
\right]^{-1}.
\end{equation}
Writing
$\rho_f\mathbb I+\rho_\Phi=(h/e^2)\left[r\,\mathbb I-2\epsilon\right]$
with $r=(e^2/h)\rho_f+2/g_\psi$,
Eq.~\eqref{eq:rho_phys_iso} gives
\begin{equation}
\label{eq:rho34_explicit}
\rho=\frac{h}{e^2}
\left[
\frac{r}{1+r^2}\mathbb I
-\frac{r^2+2}{1+r^2}\epsilon
\right].
\end{equation}
The reference tensor is not invariant under the reconstruction.
In the clean limit $r=2/g_\psi$ and the FL has $\rho=0$; the resulting tensor reproduces Eq.~\eqref{eq:intro_jump} for $\nu=-3/4$.
At finite $\rho_f$ both the critical and the Fermi-liquid reference tensors must be reconstructed with Eq.~\eqref{eq:rho_phys_iso} before taking the difference; the Fermi-liquid baseline is then nonzero. For the isotropic spinon model, with $r_f=(e^2/h)\rho_f^{\mathrm{FL}}$, the reconstructed Fermi-liquid tensor is
\begin{equation}
\rho_{\mathrm{FL}}^{\mathrm{phys}}=\frac{h}{e^2}\left[
\frac{r_f}{1+r_f^2}\mathbb I-\frac{r_f^2}{1+r_f^2}\epsilon\right].
\end{equation}
This is a model comparison at finite spinon resistivity, not a statement about the universal critical point.

The resulting values are collected in Table~\ref{tab:summary}.
The two fillings behave differently.
At $\nu=-1/2$, $\Delta\rho_{xx}=(h/e^2)/g_\psi$, so a poorly
conducting Dirac sector yields a large critical resistivity.
At $\nu=-3/4$ and $g_\psi\ll1$,
$\Delta\rho_{xx}\approx(h/e^2)g_\psi/2$ and
$\Delta\rho_{xy}\to-h/e^2$, because the filled-band Hall term dominates the reconstructed conductivity.

At zero detuning, the leading-order Dirac DC conductivity approaches a temperature-independent limit as $T/\mu\to0$ [Eqs.~\eqref{eq:sigscal} and~\eqref{eq:qbe_dimensionless}]. At fixed nonzero detuning the scaling variable $\delta/T^{1/(\nu_{\mathrm{corr}}z)}$ varies with temperature, and the measured tensor can also inherit the temperature dependence of $\rho_f$, especially at $\nu=-3/4$. The calculation concerns
clean critical transport; disorder can alter the asymptotic
critical behavior and may produce an intermediate CDW$^*$
regime~\cite{song_2024phase}.

\section{Conclusion}

In this work we have presented a large-$N$ analysis of universal electrical transport at continuous transitions between a composite Fermi liquid and a conventional metal (with a background integer quantum Hall state for $\nu=-3/4$) in moir\'e Chern bands, with explicit focus on fillings $\nu=-1/2$ and $\nu=-3/4$.
Building on the parton-based critical theories proposed in Ref.~\cite{song_2024phase}, we formulated the low-energy theory as a QED--Chern--Simons theory.
In this description, a critical charged sector realizing a bosonic Laughlin--superfluid transition is coupled to a neutral spinon Fermi surface through emergent gauge fields and Chern--Simons mixing.

A central result is the explicit derivation of an Ioffe--Larkin composition rule for the electromagnetic response.
After integrating out the matter fields to quadratic order
and solving the resulting Gaussian gauge theory, we obtained
the series combination of the spinon resistivity
and the Chern--Simons-composed charged-sector tensor.
This is the measured response at $\nu=-1/2$ and the
quarter-filled reference response at $\nu=-3/4$, where
particle--hole reconstruction restores the filled-band Hall
contribution.

To access the DC conductivity on the zero-detuning critical trajectory, where $\omega/T\to0$, we developed a large-$N$ expansion in which both the number of auxiliary replicas of the physical matter content and the Chern--Simons levels scale with $N$.
Within this framework, the collisionless Drude singularity present at $N=\infty$ is broadened by the leading $O(1/N)$ gauge-mediated collisions, which comprise inelastic scattering from the finite-frequency gauge spectrum and a quasistatic elastic channel.
Solving a quantum Boltzmann equation for the critical Dirac sector, we obtained a universal scaling function $\Sigma(\omega/T)$ and finite DC conductivities at $N=1$. After Ioffe--Larkin/Chern--Simons composition and particle--hole reconstruction, the clean-limit QCP--FL jumps are collected in Table~\ref{tab:summary}; at $\nu=-1/2$ the Hall jump is $\Delta\rho_{xy}=-h/e^2$ at this order.

A further outcome concerns the zero-frequency singularity generated by spinon Landau damping at large $\mu$. This singularity survives the Chern--Simons mixing: the Bose-weighted gauge spectrum of the mode that scatters the critical Dirac fermions retains a zero-frequency peak of finite weight that is independent of $\mu$ and $v_{\mathrm{F}}$. The spinon Fermi surface therefore adds a universal quasistatic scattering channel at this order, as at the continuous Mott transition~\cite{UniversalTransportQuantumwitczak-krempa2012}, and the large-$\mu$ limit of the gauge propagator is not uniform near zero frequency.

Our results provide transport signatures for bandwidth-tuned transitions in moir\'e materials.
In particular, the clean-limit QCP--FL resistivity jump
provides a quantitative signature of the critical regime,
while comparison at finite spinon resistivity requires
the measured-response reconstruction.
More broadly, the framework developed here parallels earlier universal-transport analyses of Mott criticality~\cite{TheoryContinuousMottsenthil2008a,UniversalTransportQuantumwitczak-krempa2012}, while extending them to a setting enriched by topological response and emergent gauge structure.
We expect that similar methods can be applied to other fractionalized quantum critical points in moir\'e systems and beyond~\cite{song2021doping,zhang2026critical}.
It would also be valuable to analyze the thermal conductivity near the critical point, along the lines of previous studies of semiconductor metal-insulator transitions~\cite{potter}.

\paragraph{Scope and limitations}
The quoted jumps are leading-order large-$N$ estimates evaluated at $N=1$. They refer to the clean limit $\rho_f\to0$, the isotropic transport limit, and the massless, particle--hole-symmetric approximation; the intraband QBE does not by itself describe the full interband optical crossover. Numerical implementation details are recorded in Appendix~\ref{app:numerics}.

\section*{Acknowledgements}

The authors are particularly thankful to T. Senthil and Haoyu Guo for suggesting this topic and helpful discussions.

\paragraph{Funding information}
XYS is supported by the Croucher Innovation Awards and the Area of Excellence scheme of the Hong Kong Research Grants Council (AoE/P-604/25-R).

\appendix

\section{Polarization functions of the Dirac sector}
\label{app:polarization}

In this appendix, we present single-variable integral forms of the
current polarization functions, which are convenient for numerical
evaluation. All results refer to the dynamical, parity-even
polarization functions of \emph{one two-cone Dirac sector}; the two
Dirac cones of a sector are included in the bubble below. Throughout
this appendix we use $v_\psi=\hbar=k_{\mathrm B}=1$.

The gauge polarization functions of the Dirac fermion $\psi$ can be
expressed in terms of the diagonal components. With the shorthand
\begin{equation}
\label{eq:Matsubara_measure}
\!\int_{\boldsymbol k,\omega_n}
:=
T\sum_{\omega_n}
\!\int\!\frac{\mathrm d^2\boldsymbol k}{(2\pi)^2},
\end{equation}
they read
\begin{equation}
\label{eq:Psi_ii_bubble}
\Psi^{ii}(\Omega_n,\boldsymbol q)
=
2\!\int_{\boldsymbol k,\omega_n}
\frac{
4\omega_n(\omega_n+\Omega_n)
}{
(\boldsymbol k^2+\omega_n^2)
[(\boldsymbol k+\boldsymbol q)^2+(\omega_n+\Omega_n)^2]
},
\end{equation}
\begin{equation}
\label{eq:Psi_00_bubble}
\Psi^{00}(\Omega_n,\boldsymbol q)
=
2\!\int_{\boldsymbol k,\omega_n}
\frac{
2[
\boldsymbol k\!\cdot\!(\boldsymbol k+\boldsymbol q)
-\omega_n(\omega_n+\Omega_n)
]
}{
(\boldsymbol k^2+\omega_n^2)
[(\boldsymbol k+\boldsymbol q)^2+(\omega_n+\Omega_n)^2]
},
\end{equation}
where $\Psi^{ii}\equiv\Psi_{xx}+\Psi_{yy}$ denotes the trace over the
two spatial components. These expressions follow from
Eq.~\eqref{eq:pi_psi} after projecting the tensor structure onto the
corresponding components. The Euclidean external three-momentum is
$(\Omega_n,\boldsymbol q)$, and $q\equiv|\boldsymbol q|$ is its
magnitude.

The Matsubara frequency sum separates the polarizations into a vacuum
part independent of temperature and a thermal part proportional to the
Fermi function. Continuing $i\Omega_n$ to $z=\Omega+i0^+$ gives the retarded
functions, which we denote by the superscript $R$.
All square roots below are taken on one and the same retarded
analytic branch. We define
\begin{equation}
\label{eq:R_analytic}
R(w,q)=\sqrt{q^2-w^2}
\end{equation}
as the function that is analytic in the upper half $w$ plane and
positive on the positive imaginary axis,
\begin{equation}
\label{eq:R_positive}
R(i\xi,q)>0,\qquad \xi>0.
\end{equation}
On the real axis this branch reads
\begin{equation}
\label{eq:retarded_branch}
R(w+i0^+,q)=
\begin{cases}
\sqrt{q^2-w^2},& |w|<q,\\[2pt]
-i\sqrt{w^2-q^2},& w>q,\\[2pt]
+i\sqrt{w^2-q^2},& w<-q.
\end{cases}
\end{equation}
We further define
\begin{equation}
\label{eq:R_branch}
z=\Omega+i0^+,\qquad
P=R(z,q),\qquad
R_\pm(k;z,q)=R(z\pm2k,q),
\end{equation}
where $P$ is the retarded root of the main text and all roots are
evaluated with the branch of Eq.~\eqref{eq:retarded_branch}. The
Fermi function is
\begin{equation}
\label{eq:nF_def}
n_{\mathrm F}(k)=\frac{1}{e^{k/T}+1}.
\end{equation}

After the Matsubara sum and the angular integration over the angle
between $\boldsymbol k$ and $\boldsymbol q$, the retarded density
component takes the form
\begin{equation}
\label{eq:Psi00_ret}
\Psi^{00,R}(z,q)
=
-\frac{q^2}{8P}
+\delta\Psi^{00,R}(z,q),
\end{equation}
where
\begin{equation}
\label{eq:dPsi00_ret}
\delta\Psi^{00,R}(z,q)
=
-\frac{2T\ln2}{\pi}
+\frac{1}{\pi P}
\!\int_0^\infty \mathrm dk\,
n_{\mathrm F}(k)
\left[
R_+(k;z,q)+R_-(k;z,q)
\right].
\end{equation}
The constant term $-2T\ln2/\pi$ in Eq.~\eqref{eq:dPsi00_ret} is a
finite thermal contribution; together with the radial integral it
produces the finite static compressibility discussed below.

The trace over the spatial components is
\begin{equation}
\label{eq:Psi_ii_ret}
\Psi^{ii,R}(z,q)
=
-\frac{P}{8}
+\frac{z^2}{8P}
+\delta\Psi^{ii,R}(z,q),
\end{equation}
where
\begin{equation}
\label{eq:dPsi_ii_ret}
\delta\Psi^{ii,R}(z,q)
=
\frac{4}{\pi P}
\!\int_0^\infty k\,\mathrm dk\,
n_{\mathrm F}(k)
\left[
\frac{k+z}{R_+(k;z,q)}
+
\frac{k-z}{R_-(k;z,q)}
\right].
\end{equation}

For $q\ne0$, the longitudinal and transverse projections of these
retarded functions are
\begin{equation}
\label{eq:Psi_L_ret}
\Psi_L^{R}(z,q)
=
\frac{P^2}{q^2}\,
\Psi^{00,R}(z,q)
\end{equation}
and
\begin{equation}
\label{eq:Psi_T_ret}
\Psi_T^{R}(z,q)
=
\Psi^{ii,R}(z,q)
+\frac{z^2}{q^2}\,
\Psi^{00,R}(z,q).
\end{equation}
The projections follow from Eq.~\eqref{eq:components} in
Appendix~\ref{app:kernels} after the analytic continuation
$i\Omega_n\to z$, which makes the coefficient of $\Psi^{00,R}$ in
$\Psi_T^{R}$ equal to $+z^2/q^2$. Because of the explicit $1/q^2$,
the homogeneous limit is taken before projection; it is given
separately in Eq.~\eqref{eq:Psi_T_q0}.

The vacuum limit is obtained by setting the thermal corrections to
zero. Equation~\eqref{eq:Psi_L_ret} gives
$\Psi_L^{\mathrm{vac}}=(P^2/q^2)(-q^2/8P)=-P/8$, while in
Eq.~\eqref{eq:Psi_T_ret} the terms $z^2/(8P)$ and
$(z^2/q^2)(-q^2/8P)$ cancel. Hence
\begin{equation}
\label{eq:Psi_vac_check}
\Psi_L^{\mathrm{vac}}
=
\Psi_T^{\mathrm{vac}}
=
-\frac{P}{8}
=
-\frac{\sqrt{q^2-z^2}}{8},
\end{equation}
which is the gauge-invariantly regulated vacuum polarization (see,
e.g., Ref.~\cite{RAO1986227}). The polarizations used in the gauge
kernels include both the vacuum and the thermal contributions.

At $z=0$ the static polarizations are
two-cone sector the static polarizations are
\begin{align}
\label{eq:l_static}
\bar\Psi_L(q)
&=
-\frac{2T\ln2}{\pi}
-\frac q8
+\frac q\pi
\!\int_0^1\mathrm du\,
\sqrt{1-u^2}\,
n_{\mathrm F}\!\left(\frac{qu}{2}\right),
\\
\label{eq:t_static}
\bar\Psi_T(q)
&=
-\frac q8
+\frac q\pi
\!\int_0^1\mathrm du\,
\frac{u^2}{\sqrt{1-u^2}}\,
n_{\mathrm F}\!\left(\frac{qu}{2}\right),
\end{align}
with $n_{\mathrm F}$ defined in Eq.~\eqref{eq:nF_def}. They follow
from the general expressions \eqref{eq:Psi_L_ret}
and~\eqref{eq:Psi_T_ret} at $z=0$: then $P=q$, and the branches give
$R_+(k;0,q)=R_-(k;0,q)=\sqrt{q^2-4k^2}$ for $k<q/2$, while the two
roots are opposite imaginary numbers for $k>q/2$, so that only the
region $k<q/2$ contributes; the substitution $u=2k/q$ then
reproduces Eqs.~\eqref{eq:l_static} and~\eqref{eq:t_static}. Their
small-$q$ and large-$q$ limits are
\begin{equation}
\label{eq:static_ward}
\begin{aligned}
\bar\Psi_L(q)
&=
-\frac{2T\ln2}{\pi}
-\frac{q^2}{24\pi T}
+O(q^4/T^3),\\
\bar\Psi_T(q)
&=
-\frac{q^2}{12\pi T}
+O(q^4/T^3),
\end{aligned}
\end{equation}
and both approach $-q/8$ for $q\gg T$. Equation
\eqref{eq:static_ward} gives the static spatial Ward identity
$\bar\Psi_T(0)=0$. The kernel value
$\bar\Psi_L(0)=-2T\ln2/\pi$ is associated with the physical
compressibility; with the present Kubo convention the positive
compressibility is $2T\ln2/\pi$. For $q\ll T$, these
static polarizations give
\begin{equation*}
 \bar{\mathcal C}(q)\simeq q^2(1+\tfrac23\ln2),
 \qquad
 \mathcal A(q)\simeq \frac{12\pi T^2}{n_\psi(1+\tfrac23\ln2)q^2}.
\end{equation*}
Here $\mathcal C$ and $\mathcal A$ are defined in
Eqs.~\eqref{eq:Cdef} and~\eqref{eq:An}.
These are the static inputs to $F_{\mathrm{el}}$ and the $z=0$
inputs to the general-frequency expressions above.

At $q=0$ the projected formulas \eqref{eq:Psi_L_ret}
and~\eqref{eq:Psi_T_ret} are singular because of the explicit
$1/q^2$; the homogeneous limit is taken before projection and is
\begin{equation}
\label{eq:Psi_T_q0}
\Psi_T^{R}(z,0)
=
\Psi_L^{R}(z,0)
=
\frac{iz}{8}
+\frac{4}{\pi}
\!\int_0^\infty \mathrm dk\,
\frac{k^2\,n_{\mathrm F}(k)}{z^2-4k^2}.
\end{equation}
Equation~\eqref{eq:Psi_T_q0} has the correct collisionless spectral
content. For $\Omega>0$ one may use
$1/(z^2-4k^2)=\mathcal P/(\Omega^2-4k^2)
-i\pi\delta(\Omega^2-4k^2)$ with
$\delta(\Omega^2-4k^2)=\delta(k-\Omega/2)/(4\Omega)$; the imaginary
part of the thermal integral is then
$-\Omega n_{\mathrm F}(\Omega/2)/4$, and Eq.~\eqref{eq:Psi_T_q0}
gives
\begin{equation}
\label{eq:Psi_T_q0_im}
\operatorname{Im}\Psi_T^{R}(\Omega,0)
=
\frac{\Omega}{8}
\tanh\!\left(\frac{\Omega}{4T}\right),
\qquad \Omega>0,
\end{equation}
and therefore
\begin{equation}
\label{eq:sigma_coh_check}
\operatorname{Re}\sigma_{\mathrm{coh}}(\Omega)
=
\frac{1}{8}
\tanh\!\left(\frac{\Omega}{4T}\right),
\end{equation}
in agreement with Eq.~\eqref{eq:re_coh} of the main text. In the
low-frequency dynamic limit, Eq.~\eqref{eq:Psi_T_q0} gives the
current-kernel limit that determines the Drude stiffness
\begin{equation}
\label{eq:Drude_stiffness}
\Psi_T^{R}(z\to0,0)
\to
-\frac{T\ln2}{\pi},
\end{equation}
which has the same $\ln2$ coefficient as the collisionless Drude
weight of Eq.~\eqref{eq:sharp} in the main text. Finally, the density
component vanishes in the homogeneous dynamic limit,
\begin{equation}
\label{eq:Psi00_q0}
\Psi^{00,R}(z,0)=0,\qquad z\ne0.
\end{equation}
For $z\ne0$ the density kernel vanishes at $q=0$,
$\Psi^{00,R}(z,0)=0$ [Eq.~\eqref{eq:Psi00_q0}], whereas the projected
current kernels remain finite:
$\Psi_L^{R}(z\to0,0)=\Psi_T^{R}(z\to0,0)=-T\ln2/\pi$
[Eq.~\eqref{eq:Drude_stiffness}]. In the opposite order of limits,
$\bar\Psi_L(q\to0)=-2T\ln2/\pi$ and $\bar\Psi_T(q\to0)=0$
[Eq.~\eqref{eq:static_ward}]. The two negative kernel values are not
the physical compressibility or Drude coefficient themselves: with the
present response-kernel convention, the positive Dirac compressibility
is $-\bar\Psi_L(0)$ and the positive Drude weight involves the negative
of the homogeneous dynamic current-kernel limit.

\begin{figure}[t]
    \centering
    \includegraphics[width=0.95\textwidth]{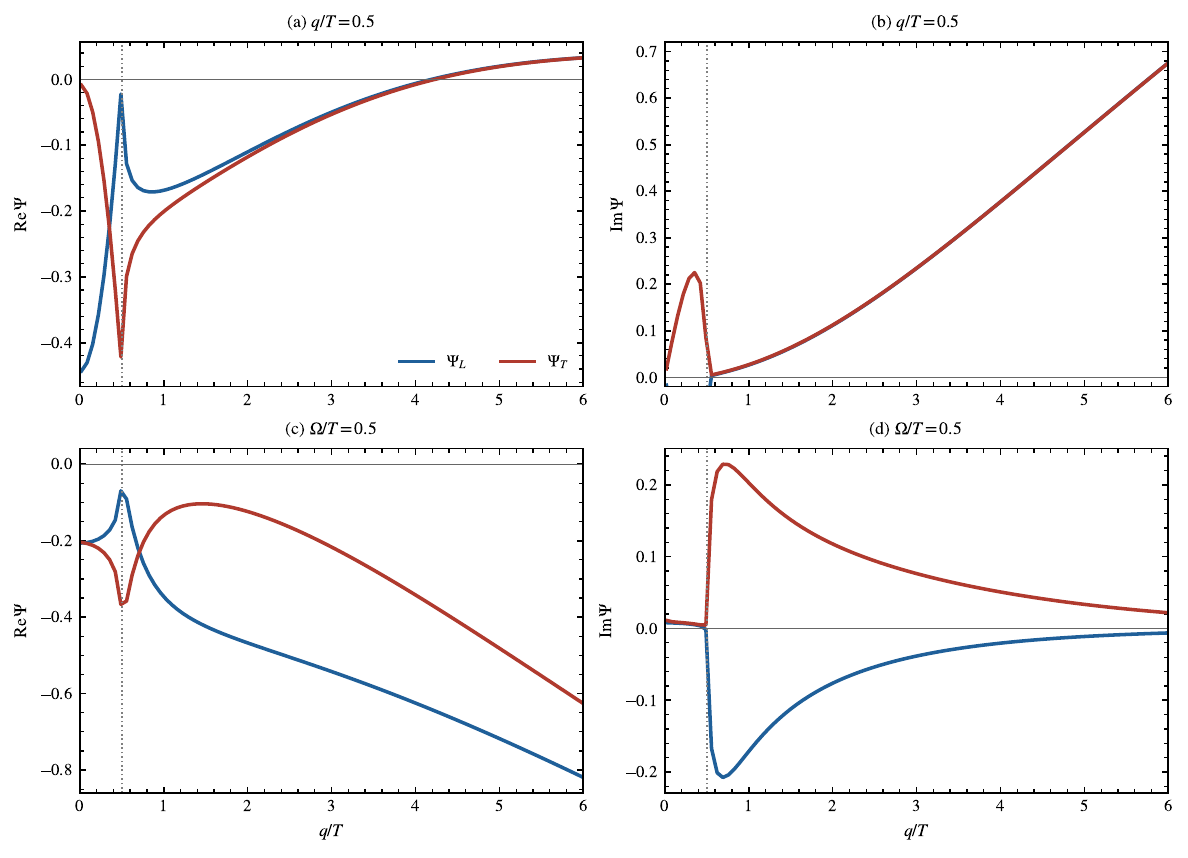}
    \caption{Real (a,c) and imaginary (b,d) parts of the longitudinal and transverse current polarization functions $\Psi_L$ and $\Psi_T$ of one two-cone Dirac sector with $T=1$ and retarded regulator $\eta/T=10^{-3}$. (a,b) show the dependence on $\Omega/T$ at fixed $q/T=0.5$; (c,d) show the dependence on $q/T$ at fixed $\Omega/T=0.5$. The dotted vertical line marks the light cone.}
    \label{fig:polarization}
\end{figure}

Figure~\ref{fig:polarization} shows representative cuts of the retarded
current polarization functions of one two-cone sector.

For numerical implementation, all roots use $z=\Omega+i\eta$.
The same fixed $\eta>0$ enters $P$ and $R_\pm$. The branch points of the radial integrals in
Eqs.~\eqref{eq:dPsi00_ret} and~\eqref{eq:dPsi_ii_ret} are located at
$k=|q\pm\Omega|/2$ and produce integrable singularities; the
remaining $k$ integrals are then well suited for direct numerical
evaluation. These radial integrals are ultraviolet finite, since
$n_{\mathrm F}(k)$ decays exponentially at large $k$ while the roots
grow at most linearly, so that the integrands vanish exponentially;
the vacuum part, by contrast, is ultraviolet divergent and is
regularized in a gauge-invariant way. The regularization concerns
only the temperature-independent piece, so the finite thermal term
$-2T\ln2/\pi$ and the compressibility it encodes are retained. The resulting functions satisfy the vacuum value
$\Psi_L^{\mathrm{vac}}=\Psi_T^{\mathrm{vac}}=-P/8$ of
Eq.~\eqref{eq:Psi_vac_check}, the static limit of
Eqs.~\eqref{eq:l_static}--\eqref{eq:static_ward}, and the homogeneous
dynamic/collisionless limits of
Eqs.~\eqref{eq:Psi_T_q0}--\eqref{eq:sigma_coh_check}. They are normalized
as the response of one two-cone sector; the two physical cones are
included in the bubble.

\section{Projected gauge-field kernels and their large-$\mu$ limits}
\label{app:kernels}

In this appendix, we illustrate the analytical methods for solving the propagator equations and express the gauge polarization function in terms of fermion bubble diagrams, Eqs.~\eqref{eq:pi_psi}, \eqref{eq:pi_fii}, and \eqref{eq:pi_f00}, for both cases.

Due to current conservation, $P_\mu\Pi^{\mu\nu}=0$, the polarization tensor admits the standard decomposition into transverse, longitudinal, and antisymmetric sectors,
\begin{equation}
\Pi^{\mu\nu}
\left(P\equiv(\Omega_n,\boldsymbol q)\right)
=
\Pi_T\mathsf P_T^{\mu\nu}
+\Pi_L\mathsf P_L^{\mu\nu}
+\Pi_O\mathsf P_O^{\mu\nu},
\end{equation}
where the projectors are defined by
\begin{gather}
\mathsf P_T^{ij}
=
\delta^{ij}
-\frac{q^iq^j}{\boldsymbol q^2},
\quad
\mathsf P_T^{0\mu}=0,
\quad
\mathsf P_O^{\mu\nu}
=
\epsilon^{\mu\nu\lambda}\frac{P^\lambda}{P},
\\
\mathsf P_L^{\mu\nu}
=
\delta^{\mu\nu}-E^{\mu\nu}-\mathsf P_T^{\mu\nu},
\qquad
E^{\mu\nu}
=
\frac{P^\mu P^\nu}{P^2}.
\end{gather}
Here $P$ is the magnitude of the Euclidean momentum $P^\mu=(\Omega_n,\boldsymbol q)$.
The tensors $\mathsf P_T^{\mu\nu}$ and $\mathsf P_L^{\mu\nu}$ project onto the
spatially transverse and the remaining gauge-invariant longitudinal subspaces.
The tensor $\mathsf P_O^{\mu\nu}$ captures the parity-odd (Hall/Chern--Simons) structure.

In this basis the Chern--Simons kernel is
$\Pi_{\mathrm{CS}}=-(P/2\pi)\mathsf P_O$.
The components can be extracted via
\begin{align}
\label{eq:components}
&\Pi_L
=
\frac{P^2}{\boldsymbol q^2}\Pi^{00},
\qquad
\Pi_T
=
\Pi^{ii}
-\frac{\Omega_n^2}{\boldsymbol q^2}\Pi^{00},
\nonumber\\
&\Pi_O
=
\frac{P}{2\Omega_n}
(\Pi^{12}-\Pi^{21}).
\end{align}
These relations follow by contracting $\Pi^{\mu\nu}$ with the corresponding projectors and using $P_\mu\Pi^{\mu\nu}=0$ to eliminate redundant components. In practice, Eq.~\eqref{eq:components} provides a convenient way to extract $\Pi_{T,L,O}$ directly from $\Pi^{00}$, $\Pi^{ii}$, and the antisymmetric spatial part.

These tensors obey the following algebra:
\begin{gather}
\mathsf P_T^2=\mathsf P_T,
\quad
\mathsf P_L^2=\mathsf P_L,
\quad
\mathsf E^2=\mathsf E,
\quad
\mathsf P_O^2=\mathsf E-\delta,
\\
\mathsf P_O\mathsf E=0,
\qquad
P_T\mathsf P_L=0,
\qquad
\mathsf P_T\mathsf E=0,
\\
\mathsf P_O\mathsf P_T+P_T\mathsf P_O=\mathsf P_O,
\qquad
P_OP_T\mathsf P_O=-\mathsf P_L.
\end{gather}

These identities encode the mutual orthogonality of the transverse and longitudinal sectors, as well as the way the parity-odd structure mixes them. They are particularly useful for algebraic manipulations of quadratic actions and for inverting kernel operators written in the $\{\mathsf P_T,\mathsf P_O,\mathsf E,\mathbb I\}$ basis, with $\mathsf P_L=\mathbb I-\mathsf E-\mathsf P_T$.

This algebra yields the following formula for the inverse of a general kernel:
\begin{equation}
\begin{split}
&(\lambda_T\mathsf P_T+\lambda_O\mathsf P_O+\lambda_E\mathsf E+\lambda_0\mathbb I)^{-1}\\
&\quad=-\frac{1}{\lambda_T\lambda_0+\lambda_O^2+\lambda_0^2}
\Big[
\lambda_T\mathsf P_T+\lambda_O\mathsf P_O
+\frac{\lambda_E\lambda_0+\lambda_E\lambda_T-\lambda_O^2}{\lambda_E+\lambda_0}\mathsf E\\
&\qquad -(\lambda_T+\lambda_0)\mathbb I
\Big].
\end{split}
\label{eq:inverse}
\end{equation}
We repeatedly use this formula to integrate out internal gauge fields and obtain the effective response kernel for the remaining mode.

The projected components $\Pi_{b,L}$, $\Pi_{b,T}$, and
$\Pi_{b,O}$ obtained below are the finite-$\mu$ kernels
of the Coulomb-gauge propagator. For the real-time analysis of
Sec.~\ref{sec:largeN} they are continued to the retarded
branch specified in Appendix~\ref{app:polarization},
with the finite-$\mu$ matter response. We write
$\Psi_{L,T}\equiv\Pi_{\psi,L,T}$ and
$F_{L,T}\equiv\Pi_{f,L,T}$ for the projected components obtained
from Eqs.~\eqref{eq:pi_psi}--\eqref{eq:pi_f00} through
Eq.~\eqref{eq:components}; both enter the kernels below on the same
footing. In the formulas below we
write $q\equiv|\boldsymbol q|$. The TT entry of the retarded
propagator, Eq.~\eqref{eq:Grets}, is
\begin{equation}
\mathcal G^R_{TT}
=
\frac{\Pi_{b,L}}{\Pi_{b,L}\Pi_{b,T}+\Pi_{b,O}^2},
\end{equation}
with all quantities evaluated on the retarded branch. The
Chern--Simons term displayed in $\Pi_{b,O}$ is the level-$N$
term left after the integer contact has cancelled the bare
$b\,\mathrm db$ term.

\subsection{Finite-$\mu$ projected kernels}

For $\nu=-1/2$, using Eq.~\eqref{eq:inverse}, Eq.~\eqref{eq:ha} gives
\begin{align}
\Pi_{b,T}
&=
\Psi_T+
\frac{P^2F_T}{P^2+4\pi^2F_LF_T},
\label{eq:pi_b1_t}\\
\Pi_{b,L}
&=
\Psi_L+
\frac{P^2F_L}{P^2+4\pi^2F_LF_T},
\label{eq:pi_b1_l}\\
\Pi_{b,O}
&=
-\frac{P}{2\pi}
+\frac{P^3}
{2\pi(P^2+4\pi^2F_LF_T)}
=
-\frac{2\pi PF_LF_T}
{P^2+4\pi^2F_LF_T}.
\label{eq:pi_b1_o}
\end{align}
The term $-P/2\pi$ in $\Pi_{b,O}$ is the level-$N$ Chern--Simons term of $b$. Physically, $\Psi_{L,T}$ encode the critical Dirac-sector response, while $F_{L,T}$ capture the contribution from the Fermi-surface sector. Equations~\eqref{eq:pi_b1_t}--\eqref{eq:pi_b1_o} are the finite-$\mu$ kernels for this filling. Substituting them into the TT entry of Eq.~\eqref{eq:Grets} gives
\begin{equation}
\label{eq:G1}
\mathcal G^R_{TT}
=
\frac{
P^2(F_L+\Psi_L)+4\pi^2\Psi_LF_LF_T
}{
P^2(F_L+\Psi_L)(F_T+\Psi_T)+4\pi^2\Psi_L\Psi_TF_LF_T
}.
\end{equation}

For $\nu=-3/4$, the effective kernel is obtained by
successively integrating out $a$, $b_2$, and $\alpha$
in the original quadratic action. On the
gauge-transverse subspace, the result can be written as
\begin{equation}
\label{eq:schur34}
\Pi_b
=\Pi_\psi
+\Pi_{\mathrm{CS}}
\Big[
2\Pi_{\mathrm{CS}}
+\Pi_{\mathrm{CS}}(\Pi_\psi^{-1}+\Pi_f^{-1})\Pi_{\mathrm{CS}}
\Big]^{-1}
\Pi_{\mathrm{CS}}.
\end{equation}
This is equivalent to Eq.~\eqref{eq:b34}.

After repeated application of Eq.~\eqref{eq:inverse}, Eq.~\eqref{eq:b34} yields

\begin{align}
\Pi_{b,T}(P)
&=
\Psi_T\,
\frac{
P^2(F_L+\Psi_L)(2F_T+\Psi_T)+16\pi^2F_LF_T\Psi_L\Psi_T
}{
\mathcal D_2
},
\label{eq:pi_b2_t}\\
\Pi_{b,L}(P)
&=
\Psi_L\,
\frac{
P^2(2F_L+\Psi_L)(F_T+\Psi_T)+16\pi^2F_LF_T\Psi_L\Psi_T
}{
\mathcal D_2
},
\label{eq:pi_b2_l}\\
\Pi_{b,O}(P)
&=
-\frac{4\pi PF_LF_T\Psi_L\Psi_T}{\mathcal D_2},
\label{eq:pi_b2_o}
\end{align}
with
$\mathcal D_2=P^2(F_L+\Psi_L)(F_T+\Psi_T)+16\pi^2F_LF_T\Psi_L\Psi_T$.
Substituting into the TT entry of Eq.~\eqref{eq:Grets},
\begin{equation}
\label{eq:G2}
\mathcal G^R_{TT}
=
\frac{
P^2(2F_L+\Psi_L)(F_T+\Psi_T)+16\pi^2F_LF_T\Psi_L\Psi_T
}{
\Psi_T\left[
P^2(2F_L+\Psi_L)(2F_T+\Psi_T)+16\pi^2F_LF_T\Psi_L\Psi_T
\right]
}.
\end{equation}

Equations~\eqref{eq:pi_b2_t}--\eqref{eq:pi_b2_o} are the finite-$\mu$
kernels for this filling, as at $\nu=-1/2$.

\subsection{Large-$\mu$ fixed-frequency limit}

At generic frequencies $\Omega\sim q,T$, the leading spinon
polarization components are proportional to the large microscopic
scale $\mu$. For $\Omega,q,T\ll\mu$ at fixed nonzero frequency,
with $P$ on the retarded branch,
$P^2=q^2-(\Omega+i0^+)^2$, Eqs.~\eqref{eq:G1}
and~\eqref{eq:G2} reduce to the transverse entry of
Eq.~\eqref{eq:Dinf} with
$n_\psi=1$ and $n_\psi=2$, respectively; the superscript
$(\infty)$ denotes the leading large-$\mu$ expression obtained at
fixed nonzero frequency. In both cases the microscopic
Fermi-surface parameters drop out of the transverse spectral
factor. The density diagonal at the same order follows from the
same kernels through
$D_0^{(\infty)}=-\operatorname{Im}\mathcal G_{00}^{(\infty)}$
of Eq.~\eqref{eq:Dinf}.

\begin{figure}[t]
    \centering
    \includegraphics[width=\textwidth]{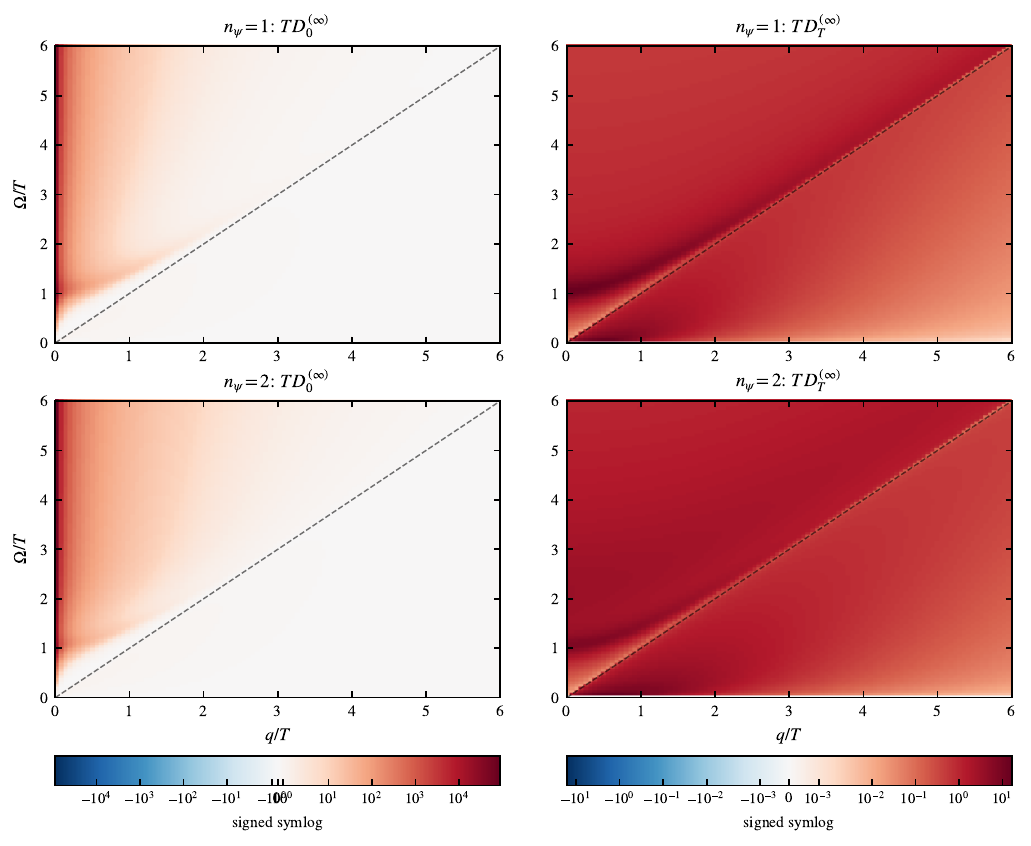}
    \caption{Regular gauge spectra $T D_0^{(\infty)}(\Omega,q)$ and $T D_T^{(\infty)}(\Omega,q)$ for $n_\psi=1$ and $n_\psi=2$. The horizontal and vertical axes are $q/T$ and $\Omega/T$; the dashed line is the light cone $\Omega=q$. The color scale is signed symlog. The heatmaps start at $q/T=0.02$; the $q=0$ limit is treated analytically in Appendix~\ref{app:polarization}.}
    \label{fig:spectra}
\end{figure}

\subsection{Boundary-layer analysis of the quasistatic channel}

The large-$\mu$ limit is not uniform near $\Omega=0$. At $T\ll\mu$, Eqs.~\eqref{eq:pi_fii} and~\eqref{eq:pi_f00} give the two-dimensional Lindhard functions for $0<\Omega<v_{\mathrm F}q$:
\begin{equation}
\label{eq:lindhard}
\left\{
\begin{aligned}
F_L(\Omega,q)
&=
\frac{\mu}{\pi}
\frac{\Omega^2-q^2}{v_{\mathrm F}^2q^2}
\left(
1+i\frac{\Omega}
{\sqrt{v_{\mathrm F}^2q^2-\Omega^2}}
\right),\\
F_T(\Omega,q)
&=
\frac{\mu}{\pi}
\left[
-\frac{\Omega^2}{v_{\mathrm{F}}^2q^2}
+i\frac{\Omega\sqrt{v_{\mathrm{F}}^2q^2-\Omega^2}}
{v_{\mathrm{F}}^2q^2}
\right].
\end{aligned}
\right.
\end{equation}
Their imaginary parts describe Landau damping by the spinon
particle--hole continuum. For $q\sim T$ they introduce the
frequency scale $T^2/\mu$. In the window $\Omega\sim T^2/\mu$,
these reduce to
$F_L\simeq-\mu/(\pi v_{\mathrm{F}}^2)$,
$F_T\simeq i\mu\Omega/(\pi v_{\mathrm{F}}q)$,
and $P^2\simeq q^2$, while the Dirac polarizations take their
static values $\bar\Psi_{L,T}(q)\equiv\Psi_{L,T}(0,q)$, whose
explicit forms are given in
Appendix~\ref{app:polarization}
[Eqs.~\eqref{eq:l_static} and~\eqref{eq:t_static}].
Substituting into the projected kernels and keeping the leading
terms in $1/\mu$ gives, on the retarded branch,
\begin{equation}
\label{eq:Dsingle}
\begin{aligned}
\mathcal G^R(\Omega,q)\simeq{}&
\begin{pmatrix}
-\dfrac{n_\psi-1}{n_\psi\bar\Psi_L} & 0\\[2mm]
0 & \dfrac{1}{\bar\Psi_T}
\end{pmatrix}\\
&{}+\frac{\mathcal A(q)}{T}\,
\frac{\Omega}{\Omega+i\Omega_*(q)}
\begin{pmatrix}
\zeta(q)^2 & i\zeta(q)\\[2mm]
-i\zeta(q) & 1
\end{pmatrix},
\quad n_\psi=1,2,
\end{aligned}
\end{equation}
where $\mathcal A(q)$, $\Omega_*(q)$, and $\zeta(q)$ are the
quantities of Eqs.~\eqref{eq:An}, \eqref{eq:Gamman},
and~\eqref{eq:c_q}. The first term is real and does not
contribute to the spectrum. Only the second term contributes, so
\begin{equation}
\frac{T}{\Omega}D_T(\Omega,q)
=
\mathcal A(q)\,
\frac{\Omega_*(q)}{\Omega^2+\Omega_*(q)^2},
\qquad
D_0=\zeta(q)^2D_T,
\end{equation}
which is the Lorentzian peak of Eq.~\eqref{eq:Wsing} with the
weight and width of Eqs.~\eqref{eq:An} and~\eqref{eq:Gamman}. Away
from the peak, $\Omega/\Omega_*\to\infty$, the propagator
approaches the static limit of the fixed-frequency expression, so
the regular and quasistatic parts are not double counted.

The same boundary-layer analysis applies at $\nu=-3/4$. With the
same replacements and $n_\psi=2$, the TT entry of
Eq.~\eqref{eq:Grets} reduces to Eq.~\eqref{eq:Dsingle}; compared
with $\nu=-1/2$ at the same static polarizations, the peak has the
same height, half the width, and half the weight.

Together with the regular spectra of Eq.~\eqref{eq:Dinf},
Eq.~\eqref{eq:Dsingle} defines the gauge spectrum entering the
collision integral of Sec.~\ref{sec:qbe}.

\section{Linearized collision integral}
\label{app:collision}

In this appendix, we present the details for linearizing the scattering terms of Eq.~\eqref{eq:qbe}, which are built from the Coulomb-gauge density/transverse-current propagator. The collision integral consists of a regular part, built from the fixed-frequency diagonal spectra $D_{0,T}^{(\infty)}$ of Eq.~\eqref{eq:Dinf}, and a quasistatic part generated by the narrow peak of Sec.~\ref{sec:gauge} (Appendix~\ref{app:kernels}). We linearize them in turn. Our goal is to rewrite the collision integral in a form suitable for the linearized QBE, i.e. as a linear integral acting on the nonequilibrium distribution $\delta f$, or equivalently on the scalar function $\varphi$ introduced in the main text.

We use $\boldsymbol k'=\boldsymbol k+\boldsymbol q$ and write
$\delta f(\boldsymbol k)=\delta f_+(\boldsymbol k)$.
Particle--hole symmetry in linear response gives
$\delta f_-(\boldsymbol k)=-\delta f(\boldsymbol k)$.

Starting from Eq.~\eqref{eq:qbe}, retaining the linear order in $\delta f$, and integrating over the frequency $\Omega$, we obtain
\begin{equation}
\label{eq:I_ha}
\begin{aligned}
I_b^{\mathrm{reg}}
={}&
-\frac{2}{N}
\!\int\!\frac{\mathrm d^2q}{(2\pi)^2}\\
&\times
\big\{
W_+(k,k',\theta)
[
|\vartheta_+(k,k')|\delta f(k)
-|\vartheta_+(k',k)|\delta f(k')
]\\
&\quad+
W_-(k,k',\theta)
[
\vartheta_-(k,k')\delta f(k)
+\vartheta_-(k',k)\delta f(k')
]
\big\},
\end{aligned}
\end{equation}
where $W_\pm(k,k',\theta)$ are the on-shell weights of Eq.~\eqref{eq:Wpm}, with $\Omega=|k-k'|$ for $W_+$ and $\Omega=k+k'$ for $W_-$.
The two terms in braces correspond to two distinct kinematic channels: the first involves energy transfer $|\varepsilon_k-\epsilon_{k'}|$ and describes scattering with absorption or emission of a gauge boson, while the second involves $|\varepsilon_k+\epsilon_{k'}|$ and describes particle--hole pair processes.

The two functions are defined by
\begin{gather}
\vartheta_+(k,k')
=
\frac{
n_{\mathrm F}(k')[1-n_{\mathrm F}(k')]
}{
n_{\mathrm F}(k')-n_{\mathrm F}(k)
},
\\
\vartheta_-(k,k')
=
\frac{
[1-n_{\mathrm F}(k')]n_{\mathrm F}(k')
}{
1-n_{\mathrm F}(k)-n_{\mathrm F}(k')
}.
\end{gather}
They correspond to the absorption/emission channel and the particle--hole pair channel, respectively, and both follow from the frequency integral in Eq.~\eqref{eq:qbe} after collecting the remaining equilibrium distribution factors. In particular, $\vartheta_+$ is associated with the gain--loss structure for transitions between the single-particle states $k\to k'$, while $\vartheta_-$ encodes the phase-space factor appropriate for the pair channel.

The quasistatic part follows from Eq.~\eqref{eq:Iel}. Inserting Eq.~\eqref{eq:fslinear} and using the elastic-shell contraction, for which the on-shell weight is the density-plus-transverse form of Eq.~\eqref{eq:Iel}, and averaging over the angle $\theta$ between $\boldsymbol k$ and $\boldsymbol k'$,
\begin{equation}
\label{eq:Iel_lin}
I_{\mathrm{el}}
=-\frac{T}{N}\,\boldsymbol E\!\cdot\!\boldsymbol k\,
F_{\mathrm{el}}\!\left(\frac{k}{T}\right)\varphi(k),
\end{equation}
with $F_{\mathrm{el}}$ given in Eq.~\eqref{eq:Fel}. For $q\ll T$, $\mathcal A(q)\propto T^2/q^2$, while the density bracket $1+\zeta(q)^2\cos^2(\theta/2)$ tends to unity. Elastic out-scattering and in-scattering therefore diverge separately in the forward direction and must be combined with a common infrared cutoff before the cutoff is removed; the combination $(1-\cos\theta)\mathcal A(q)[1+\zeta(q)^2\cos^2(\theta/2)]$ is finite because $q^2=2k^2(1-\cos\theta)$. The same ordering is used for the regular part.

Inserting Eq.~\eqref{eq:fslinear} into Eq.~\eqref{eq:I_ha} and adding Eq.~\eqref{eq:Iel_lin}, the collision term takes the form
\begin{equation}
I_b
=
\frac{T}{N}\boldsymbol E\!\cdot\!\boldsymbol k
\Big[
-\Big(F_b\!\left(\frac{k}{T}\right)
+F_{\mathrm{el}}\!\left(\frac{k}{T}\right)\Big)\varphi(k)
+\!\int\!\frac{\mathrm dk'}{T}
K_b(k'/T,k/T)\varphi(k')
\Big].
\end{equation}
In this representation, the collision term is explicitly linear in the driving field $\boldsymbol E$ and in the unknown function $\varphi(k)$, and it is separated into an out-scattering part proportional to $-F_b\varphi(k)$ and an in-scattering part represented by the integral kernel $K_b(k',k)$. For the regular part, the split into $F_b$ and $K_b$ is not a decomposition into elastic and inelastic scattering; the elastic quasistatic channel is contained entirely in $F_{\mathrm{el}}$. The overall factor $T/N$ is the same large-$N$ normalization as in Eq.~\eqref{eq:qbe}.

In the following definitions of the dimensionless kernels, momenta and frequencies are measured in units of $T$, and the regular spectra are correspondingly evaluated at $T=1$. The functions $F_b$ and $K_b$ then read
\begin{equation}
F_b(k)
=
\!\int\!\frac{k'\mathrm dk'}{\pi}
\Big[
\langle W_+\rangle_\theta\,|\vartheta_+(k,k')|
+\langle W_-\rangle_\theta\,\vartheta_-(k,k')
\Big],
\end{equation}
\begin{equation}
K_b(k',k)
=
\frac{k'^2}{\pi k}
\Big[
\langle W_+\cos\theta\rangle_\theta\,|\vartheta_+(k',k)|
-\langle W_-\cos\theta\rangle_\theta\,\vartheta_-(k',k)
\Big].
\end{equation}
Here the angular dependence of the scattering has been integrated out and absorbed into the angular average
\begin{equation}
\langle X\rangle_\theta
=
\!\int_0^{2\pi}\frac{\mathrm d\theta}{2\pi}\,X,
\end{equation}
with $q=|\boldsymbol k'-\boldsymbol k|=\sqrt{k^2+k'^2-2kk'\cos\theta}$, where $\theta$ is the angle between $\boldsymbol k$ and $\boldsymbol k'$. $F_b$ collects all contributions in which the distribution distortion is evaluated at the same momentum $k$, whereas $K_b$ multiplies $\varphi(k')$ and describes scattering into the state $k$ from other states $k'$. The factors of $k'/k$ and $\cos\theta$ in $K_b$ arise from projecting the distribution distortion onto $\boldsymbol E\!\cdot\!\boldsymbol k$.

\section{Numerical solution of the QBE}
\label{app:numerics}

This appendix records the numerical parameters used for the results reported in the main text.

\subsection{Radial grid and ultraviolet cutoff}
The fermion momentum grid consists of $N_k=28$ Gauss--Legendre nodes on $[0,\Lambda]$ with $\Lambda/T=16$. The angular average uses $N_\theta=24$ uniformly spaced nodes on $[0,2\pi)$. The polarization integrals of Appendix~\ref{app:polarization} use $N_k^{(\Psi)}=64$ Gauss--Legendre nodes in each interval obtained by splitting at the branch points $k=|q\pm\Omega|/2$, with a thermal cutoff $k_{\max}/T=20$.

\subsection{Common infrared regulator}
The scattering momenta are evaluated with the same floor $q/T\ge 2\times10^{-4}$ in the out- and in-scattering terms. No separate cutoff is used for the two terms, and the polarization regulator is shared by both collision channels.

\subsection{Regulator extrapolation}
The retarded polarization uses $\eta/T=2\times10^{-4}$ for the production run. The regulator dependence is monitored through the refinement sequence reported below; the final values are stable at the quoted numerical accuracy.

\subsection{Linear solver and convergence}
At $N=1$, the linearized collision operator is a dense $28\times28$ matrix and is solved directly for each frequency. The frequency grid contains the DC point and 80 points in $0.02\le\omega/T\le8$. The four refinement sets give the following values of $g_\psi$.
\begin{center}
\begin{tabular}{@{}rrrrcc@{}}
\hline
$N_k$ & $N_\theta$ & $\Lambda/T$ & $\eta/T$ & $g_\psi$ ($n_\psi=1$) & $g_\psi$ ($n_\psi=2$) \\
\hline
16 & 12 & 10 & $10^{-3}$        & 0.170982 & 0.266823 \\
20 & 16 & 12 & $10^{-3}$        & 0.170527 & 0.263814 \\
24 & 20 & 14 & $5\times10^{-4}$ & 0.170017 & 0.261225 \\
28 & 24 & 16 & $2\times10^{-4}$ & 0.169599 & 0.259086 \\
\hline
\end{tabular}
\end{center}
The $g_\psi$ values from the last two sets differ by about $0.3\%$ and
$0.8\%$, respectively.

\nolinenumbers
\bibliography{references}

@article{zhou2024fractional,
  title = {Fractional Quantum Anomalous Hall Effect in Rhombohedral Multilayer Graphene in the Moir\'eless Limit},
  author = {Zhou, Boran and Yang, Hui and Zhang, Ya-Hui},
  journal = {Phys. Rev. Lett.},
  volume = {133},
  issue = {20},
  pages = {206504},
  numpages = {7},
  year = {2024},
  month = {Nov},
  publisher = {American Physical Society},
  doi = {10.1103/PhysRevLett.133.206504},
  url = {https://link.aps.org/doi/10.1103/PhysRevLett.133.206504}
}

@article{zhang2026critical,
  title = {Phases and criticality of the triangular lattice $\mathrm{SU}(N)$ Hofstadter-Hubbard model},
  author = {Zhang, Lu and Liu, Rongning and Song, Xue-Yang},
  journal = {Phys. Rev. B},
  volume = {113},
  issue = {4},
  pages = {045131},
  numpages = {11},
  year = {2026},
  month = {Jan},
  publisher = {American Physical Society},
  doi = {10.1103/j9mv-75ls},
  url = {https://link.aps.org/doi/10.1103/j9mv-75ls}
}

@article{song2021doping,
  title = {Doping the chiral spin liquid: Topological superconductor or chiral metal},
  author = {Song, Xue-Yang and Vishwanath, Ashvin and Zhang, Ya-Hui},
  journal = {Phys. Rev. B},
  volume = {103},
  issue = {16},
  pages = {165138},
  numpages = {16},
  year = {2021},
  month = {Apr},
  publisher = {American Physical Society},
  doi = {10.1103/PhysRevB.103.165138},
  url = {https://link.aps.org/doi/10.1103/PhysRevB.103.165138}
}

@article{dong2024theory,
  title = {Theory of Quantum Anomalous Hall Phases in Pentalayer Rhombohedral Graphene Moir\'e Structures},
  author = {Dong, Zhihuan and Patri, Adarsh S. and Senthil, T.},
  journal = {Phys. Rev. Lett.},
  volume = {133},
  issue = {20},
  pages = {206502},
  numpages = {6},
  year = {2024},
  month = {Nov},
  publisher = {American Physical Society},
  doi = {10.1103/PhysRevLett.133.206502},
  url = {https://link.aps.org/doi/10.1103/PhysRevLett.133.206502}
}

@article{dong2024anomalous,
  title = {Anomalous Hall Crystals in Rhombohedral Multilayer Graphene. I. Interaction-Driven Chern Bands and Fractional Quantum Hall States at Zero Magnetic Field},
  author = {Dong, Junkai and Wang, Taige and Wang, Tianle and Soejima, Tomohiro and Zaletel, Michael P. and Vishwanath, Ashvin and Parker, Daniel E.},
  journal = {Phys. Rev. Lett.},
  volume = {133},
  issue = {20},
  pages = {206503},
  numpages = {11},
  year = {2024},
  month = {Nov},
  publisher = {American Physical Society},
  doi = {10.1103/PhysRevLett.133.206503},
  url = {https://link.aps.org/doi/10.1103/PhysRevLett.133.206503}
}

@article{lu2023fractional,
  author = {Lu, Zhengguang and Han, Tonghang and Yao, Yuxuan and Reddy, Aidan P. and Yang, Jixiang and Seo, Junseok and Watanabe, Kenji and Taniguchi, Takashi and Fu, Liang and Ju, Long},
  title = {Fractional quantum anomalous Hall effect in multilayer graphene},
  journal = {Nature},
  year = {2024},
  volume = {626},
  pages = {759--764}
}

@article{nuckolls2020strongly,
  title={Strongly correlated Chern insulators in magic-angle twisted bilayer graphene},
  author={Nuckolls, Kevin P and Oh, Myungchul and Wong, Dillon and Lian, Biao and Watanabe, Kenji and Taniguchi, Takashi and Bernevig, B Andrei and Yazdani, Ali},
  journal={Nature},
  volume={588},
  number={7839},
  pages={610--615},
  year={2020},
  publisher={Nature Publishing Group UK London}
}

@article{TheoryContinuousMottsenthil2008a,
	title = {Theory of a continuous {Mott} transition in two dimensions},
	volume = {78},
	url = {https://link.aps.org/doi/10.1103/PhysRevB.78.045109},
	doi = {10.1103/PhysRevB.78.045109},
	number = {4},
	urldate = {2024-12-16},
	journal = {Physical Review B},
	author = {Senthil, T.},
	year = {2008},
	pages = {045109},
}

@article{WeakMagnetismNonFermisenthil2004,
	title = {Weak magnetism and non-{Fermi} liquids near heavy-fermion critical points},
	volume = {69},
	url = {https://link.aps.org/doi/10.1103/PhysRevB.69.035111},
	doi = {10.1103/PhysRevB.69.035111},
	number = {3},
	urldate = {2024-12-13},
	journal = {Physical Review B},
	author = {Senthil, T. and Vojta, Matthias and Sachdev, Subir},
	month = jan,
	year = {2004},
	pages = {035111},

}

@article{NonzerotemperatureTransportFractionalsachdev1998a,
	title = {Nonzero-temperature transport near fractional quantum {Hall} critical points},
	volume = {57},
	url = {https://link.aps.org/doi/10.1103/PhysRevB.57.7157},
	doi = {10.1103/PhysRevB.57.7157},
	number = {12},
	urldate = {2024-12-05},
	journal = {Physical Review B},
	author = {Sachdev, Subir},
	month = mar,
	year = {1998},
	pages = {7157--7173},
}

@article{EmergentMultiflavorQED3lee2018,
	title = {Emergent {Multi}-flavor {QED3} at the {Plateau} {Transition} between {Fractional} {Chern} {Insulators}: {Applications} to graphene heterostructures},
	volume = {8},
	issn = {2160-3308},
	shorttitle = {Emergent {Multi}-flavor {QED3} at the {Plateau} {Transition} between {Fractional} {Chern} {Insulators}},
	url = {http://arxiv.org/abs/1802.09538},
	doi = {10.1103/PhysRevX.8.031015},
	number = {3},
	urldate = {2025-06-17},
	journal = {Physical Review X},
	author = {Lee, Jong Yeon and Wang, Chong and Zaletel, Michael P. and Vishwanath, Ashvin and He, Yin-Chen},
	month = jul,
	year = {2018},
	note = {arXiv:1802.09538 [cond-mat]},
	pages = {031015},
}

@article{GaplessFermionsGaugeioffe1989,
	title = {Gapless fermions and gauge fields in dielectrics},
	volume = {39},
	url = {https://link.aps.org/doi/10.1103/PhysRevB.39.8988},
	doi = {10.1103/PhysRevB.39.8988},
	number = {13},
	urldate = {2024-12-13},
	journal = {Physical Review B},
	author = {Ioffe, L. B. and Larkin, A. I.},
	month = may,
	year = {1989},
	pages = {8988--8999},
}

@article{UniversalTransportQuantumwitczak-krempa2012,
	title = {Universal transport near a quantum critical {Mott} transition in two dimensions},
	volume = {86},
	url = {https://link.aps.org/doi/10.1103/PhysRevB.86.245102},
	doi = {10.1103/PhysRevB.86.245102},
	number = {24},
	urldate = {2024-12-05},
	journal = {Physical Review B},
	author = {Witczak-Krempa, William and Ghaemi, Pouyan and Senthil, T. and Kim, Yong Baek},
	month = dec,
	year = {2012},
	pages = {245102},
}

@article{TransitionsQuantumHallwen1993,
	title = {Transitions between the quantum {Hall} states and insulators induced by periodic potentials},
	volume = {70},
	url = {https://link.aps.org/doi/10.1103/PhysRevLett.70.1501},
	doi = {10.1103/PhysRevLett.70.1501},
	number = {10},
	urldate = {2025-02-20},
	journal = {Physical Review Letters},
	author = {Wen, Xiao-Gang and Wu, Yong-Shi},
	month = mar,
	year = {1993},
	pages = {1501--1504},
}

@article{MottTransitionAnyonchen1993,
	title = {Mott transition in an anyon gas},
	volume = {48},
	url = {https://link.aps.org/doi/10.1103/PhysRevB.48.13749},
	doi = {10.1103/PhysRevB.48.13749},
	number = {18},
	urldate = {2025-02-20},
	journal = {Physical Review B},
	author = {Chen, Wei and Fisher, Matthew P. A. and Wu, Yong-Shi},
	month = nov,
	year = {1993},
	pages = {13749--13761},
}

@article{QuantumCriticalityU1kaul2008,
	title = {Quantum criticality of {U}(1) gauge theories with fermionic and bosonic matter in two spatial dimensions},
	volume = {77},
	url = {https://link.aps.org/doi/10.1103/PhysRevB.77.155105},
	doi = {10.1103/PhysRevB.77.155105},
	number = {15},
	urldate = {2025-02-20},
	journal = {Physical Review B},
	author = {Kaul, Ribhu K. and Sachdev, Subir},
	month = apr,
	year = {2008},
	pages = {155105},
}

@article{ZeroFieldCompositeFermigoldman2023,
	title = {Zero-{Field} {Composite} {Fermi} {Liquid} in {Twisted} {Semiconductor} {Bilayers}},
	volume = {131},
	url = {https://link.aps.org/doi/10.1103/PhysRevLett.131.136501},
	doi = {10.1103/PhysRevLett.131.136501},
	number = {13},
	urldate = {2025-02-19},
	journal = {Physical Review Letters},
	author = {Goldman, Hart and Reddy, Aidan P. and Paul, Nisarga and Fu, Liang},
	month = sep,
	year = {2023},
	pages = {136501},
}

@article{CompositeFermiLiquiddong2023,
	title = {Composite {Fermi} {Liquid} at {Zero} {Magnetic} {Field} in {Twisted} \$\{{\textbackslash}mathrm\{{MoTe}\}\}\_\{2\}\$},
	volume = {131},
	url = {https://link.aps.org/doi/10.1103/PhysRevLett.131.136502},
	doi = {10.1103/PhysRevLett.131.136502},
	number = {13},
	urldate = {2025-02-19},
	journal = {Physical Review Letters},
	author = {Dong, Junkai and Wang, Jie and Ledwith, Patrick J. and Vishwanath, Ashvin and Parker, Daniel E.},
	month = sep,
	year = {2023},
	pages = {136502},
}

@article{QuantumCriticalTransportfritz2011,
	title = {Quantum critical transport at a semimetal-to-insulator transition on the honeycomb lattice},
	volume = {83},
	url = {https://link.aps.org/doi/10.1103/PhysRevB.83.035125},
	doi = {10.1103/PhysRevB.83.035125},
	number = {3},
	urldate = {2025-02-17},
	journal = {Physical Review B},
	author = {Fritz, Lars},
	month = jan,
	year = {2011},
	pages = {035125},
}

@article{QuantumCriticalTransportfritz2008a,
	title = {Quantum critical transport in clean graphene},
	volume = {78},
	url = {https://link.aps.org/doi/10.1103/PhysRevB.78.085416},
	doi = {10.1103/PhysRevB.78.085416},
	number = {8},
	urldate = {2025-02-17},
	journal = {Physical Review B},
	author = {Fritz, Lars and Schmalian, Jörg and Müller, Markus and Sachdev, Subir},
	month = aug,
	year = {2008},
	pages = {085416},
}

@article{NonzerotemperatureTransportQuantumdamle1997,
	title = {Nonzero-temperature transport near quantum critical points},
	volume = {56},
	url = {https://link.aps.org/doi/10.1103/PhysRevB.56.8714},
	doi = {10.1103/PhysRevB.56.8714},
	number = {14},
	urldate = {2025-02-17},
	journal = {Physical Review B},
	author = {Damle, Kedar and Sachdev, Subir},
	month = oct,
	year = {1997},
	pages = {8714--8733},
}

@article{ContinuousTransitionsCompositebarkeshli2012,
	title = {Continuous transitions between composite {Fermi} liquid and {Landau} {Fermi} liquid: {A} route to fractionalized {Mott} insulators},
	volume = {86},
	shorttitle = {Continuous transitions between composite {Fermi} liquid and {Landau} {Fermi} liquid},
	url = {https://link.aps.org/doi/10.1103/PhysRevB.86.075136},
	doi = {10.1103/PhysRevB.86.075136},
	number = {7},
	urldate = {2025-02-07},
	journal = {Physical Review B},
	author = {Barkeshli, Maissam and McGreevy, John},
	month = aug,
	year = {2012},
	pages = {075136},
}

@article{RAO1986227,
  title = {Parity Anomalies in Gauge Theories in 2 + 1 Dimensions},
  author = {Rao, Sumathi and Yahalom, Ram},
  year = 1986,
  journal = {Physics Letters B},
  volume = {172},
  number = {2},
  pages = {227--230},
  issn = {0370-2693},
  doi = {10.1016/0370-2693(86)90840-3}
}

@article{caiSignaturesFractionalQuantum2023,
  title = {Signatures of Fractional Quantum Anomalous {{Hall}} States in Twisted {{MoTe2}}},
  author = {Cai, Jiaqi and Anderson, Eric and Wang, Chong and Zhang, Xiaowei and Liu, Xiaoyu and Holtzmann, William and Zhang, Yinong and Fan, Fengren and Taniguchi, Takashi and Watanabe, Kenji and Ran, Ying and Cao, Ting and Fu, Liang and Xiao, Di and Yao, Wang and Xu, Xiaodong},
  date = {2023-10-01},
  journal = {Nature},
  shortjournal = {Nature},
  year={2023},
  volume = {622},
  number = {7981},
  pages = {63--68},
  issn = {1476-4687},
  doi = {10.1038/s41586-023-06289-w},
  url = {https://doi.org/10.1038/s41586-023-06289-w}
}

@article{zengThermodynamicEvidenceFractional2023,
  title = {Thermodynamic Evidence of Fractional {{Chern}} Insulator in Moiré {{MoTe2}}},
  author = {Zeng, Yihang and Xia, Zhengchao and Kang, Kaifei and Zhu, Jiacheng and Knüppel, Patrick and Vaswani, Chirag and Watanabe, Kenji and Taniguchi, Takashi and Mak, Kin Fai and Shan, Jie},
  date = {2023-10-01},
  year={2023},
  journal = {Nature},
  shortjournal = {Nature},
  volume = {622},
  number = {7981},
  pages = {69--73},
  issn = {1476-4687},
  doi = {10.1038/s41586-023-06452-3},
  url = {https://doi.org/10.1038/s41586-023-06452-3}
}

@article{Foutty_2024,
   title={Mapping twist-tuned multiband topology in bilayer WSe
            2},
   volume={384},
   ISSN={1095-9203},
   url={http://dx.doi.org/10.1126/science.adi4728},
   DOI={10.1126/science.adi4728},
   number={6693},
   journal={Science},
   publisher={American Association for the Advancement of Science (AAAS)},
   author={Foutty, Benjamin A. and Kometter, Carlos R. and Devakul, Trithep and Reddy, Aidan P. and Watanabe, Kenji and Taniguchi, Takashi and Fu, Liang and Feldman, Benjamin E.},
   year={2024},
   month=apr, pages={343–347} }

@article{liQuantumAnomalousHall2021,
  title = {Quantum Anomalous {{Hall}} Effect from Intertwined Moiré Bands},
  author = {Li, Tingxin and Jiang, Shengwei and Shen, Bowen and Zhang, Yang and Li, Lizhong and Tao, Zui and Devakul, Trithep and Watanabe, Kenji and Taniguchi, Takashi and Fu, Liang and Shan, Jie and Mak, Kin Fai},
  year={2021},
  date = {2021-12-01},
  journal= {Nature},
  shortjournal = {Nature},
  volume = {600},
  number = {7890},
  pages = {641--646},
  issn = {1476-4687},
  doi = {10.1038/s41586-021-04171-1},
  url = {https://doi.org/10.1038/s41586-021-04171-1}
}

@article{PhysRevB.89.235116,
  title = {Continuous transition between fractional quantum Hall and superfluid states},
  author = {Barkeshli, Maissam and McGreevy, John},
  journal = {Phys. Rev. B},
  volume = {89},
  issue = {23},
  pages = {235116},
  numpages = {6},
  year = {2014},
  month = {Jun},
  publisher = {American Physical Society},
  doi = {10.1103/PhysRevB.89.235116},
  url = {https://link.aps.org/doi/10.1103/PhysRevB.89.235116}
}

@article{song_2024phase,
  title = {Phase transitions out of quantum Hall states in moir\'e materials},
  author = {Song, Xue-Yang and Zhang, Ya-Hui and Senthil, T.},
  journal = {Phys. Rev. B},
  volume = {109},
  issue = {8},
  pages = {085143},
  numpages = {20},
  year = {2024},
  month = {Feb},
  publisher = {American Physical Society},
  doi = {10.1103/PhysRevB.109.085143},
  url = {https://link.aps.org/doi/10.1103/PhysRevB.109.085143}
}

@article{song_2024_intertwined,
  title = {Intertwined fractional quantum anomalous Hall states and charge density waves},
  author = {Song, Xue-Yang and Jian, Chao-Ming and Fu, Liang and Xu, Cenke},
  journal = {Phys. Rev. B},
  volume = {109},
  issue = {11},
  pages = {115116},
  numpages = {9},
  year = {2024},
  month = {Mar},
  publisher = {American Physical Society},
  doi = {10.1103/PhysRevB.109.115116},
  url = {https://link.aps.org/doi/10.1103/PhysRevB.109.115116}
}

@article{shi2025doping,
  title = {Doping a Fractional Quantum Anomalous Hall Insulator},
  author = {Shi, Zhengyan Darius and Senthil, T.},
  journal = {Phys. Rev. X},
  volume = {15},
  issue = {3},
  pages = {031069},
  numpages = {31},
  year = {2025},
  month = {Sep},
  publisher = {American Physical Society},
  doi = {10.1103/kcm5-hx56},
  url = {https://link.aps.org/doi/10.1103/kcm5-hx56}
}

@article{kim2025topological,
  title = {Topological chiral superconductivity beyond pairing in a Fermi liquid},
  author = {Kim, Minho and Timmel, Abigail and Ju, Long and Wen, Xiao-Gang},
  journal = {Phys. Rev. B},
  volume = {111},
  issue = {1},
  pages = {014508},
  numpages = {20},
  year = {2025},
  month = {Jan},
  publisher = {American Physical Society},
  doi = {10.1103/PhysRevB.111.014508},
  url = {https://link.aps.org/doi/10.1103/PhysRevB.111.014508}
}

@article{zhang2019nearly,
  title={Nearly flat Chern bands in moir{\'e} superlattices},
  author={Zhang, Ya-Hui and Mao, Dan and Cao, Yuan and Jarillo-Herrero, Pablo and Senthil, T},
  journal={Physical Review B},
  volume={99},
  number={7},
  pages={075127},
  year={2019},
  publisher={APS}
}

@article{repellin2020chern,
  title={Chern bands of twisted bilayer graphene: Fractional Chern insulators and spin phase transition},
  author={Repellin, C{\'e}cile and Senthil, T},
  journal={Physical Review Research},
  volume={2},
  number={2},
  pages={023238},
  year={2020},
  publisher={APS}
}

@article{xie2021fractional,
  title={Fractional Chern insulators in magic-angle twisted bilayer graphene},
  author={Xie, Yonglong and Pierce, Andrew T and Park, Jeong Min and Parker, Daniel E and Khalaf, Eslam and Ledwith, Patrick and Cao, Yuan and Lee, Seung Hwan and Chen, Shaowen and Forrester, Patrick R and others},
  journal={Nature},
  volume={600},
  number={7889},
  pages={439--443},
  year={2021},
  publisher={Nature Publishing Group UK London}
}

@article{abouelkomsan2020particle,
  title={Particle-hole duality, emergent fermi liquids, and fractional chern insulators in moir{\'e} flatbands},
  author={Abouelkomsan, Ahmed and Liu, Zhao and Bergholtz, Emil J},
  journal={Physical review letters},
  volume={124},
  number={10},
  pages={106803},
  year={2020},
  publisher={APS}
}

@article{serlin2020intrinsic,
  title={Intrinsic quantized anomalous Hall effect in a moir{\'e} heterostructure},
  author={Serlin, M and Tschirhart, CL and Polshyn, H and Zhang, Y and Zhu, J and Watanabe, K and Taniguchi, T and Balents, L and Young, AF},
  journal={Science},
  volume={367},
  number={6480},
  pages={900--903},
  year={2020},
  publisher={American Association for the Advancement of Science}
}

@article{sharpe2019emergent,
  title={Emergent ferromagnetism near three-quarters filling in twisted bilayer graphene},
  author={Sharpe, Aaron L and Fox, Eli J and Barnard, Arthur W and Finney, Joe and Watanabe, Kenji and Taniguchi, Takashi and Kastner, MA and Goldhaber-Gordon, David},
  journal={Science},
  volume={365},
  number={6453},
  pages={605--608},
  year={2019},
  publisher={American Association for the Advancement of Science}
}

@article{zhang2024moore,
  title = {Moore-Read state in half-filled moir\'e Chern band from three-body pseudopotential},
  author = {Zhang, Lu and Song, Xue-Yang},
  journal = {Phys. Rev. B},
  volume = {109},
  issue = {24},
  pages = {245128},
  numpages = {9},
  year = {2024},
  month = {Jun},
  publisher = {American Physical Society},
  doi = {10.1103/PhysRevB.109.245128},
  url = {https://link.aps.org/doi/10.1103/PhysRevB.109.245128}
}

@article{luiz2,
  title = {Fractional Chern Insulators in Twisted Bilayer ${\mathrm{MoTe}}_{2}$: A Composite Fermion Perspective},
  author = {Lu, Tianhong and Santos, Luiz H.},
  journal = {Phys. Rev. Lett.},
  volume = {133},
  issue = {18},
  pages = {186602},
  numpages = {6},
  year = {2024},
  month = {Oct},
  publisher = {American Physical Society},
  doi = {10.1103/PhysRevLett.133.186602},
  url = {https://link.aps.org/doi/10.1103/PhysRevLett.133.186602}
}

@article{luiz1,
  title = {Electromagnetic response and emergent topological orders in transition metal dichalcogenide ${\mathrm{MoTe}}_{2}$ bilayers},
  author = {Lu, Tianhong and Wu, Yi-Ming and Santos, Luiz H.},
  journal = {Phys. Rev. B},
  volume = {112},
  issue = {8},
  pages = {085138},
  numpages = {14},
  year = {2025},
  month = {Aug},
  publisher = {American Physical Society},
  doi = {10.1103/3crp-gb3d},
  url = {https://link.aps.org/doi/10.1103/3crp-gb3d}
}

@article{park2023observation,
	Author = {Park, Heonjoon and Cai, Jiaqi and Anderson, Eric and Zhang, Yinong and Zhu, Jiayi and Liu, Xiaoyu and Wang, Chong and Holtzmann, William and Hu, Chaowei and Liu, Zhaoyu and Taniguchi, Takashi and Watanabe, Kenji and Chu, Jiun-Haw and Cao, Ting and Fu, Liang and Yao, Wang and Chang, Cui-Zu and Cobden, David and Xiao, Di and Xu, Xiaodong},
	Da = {2023/10/01},
	Doi = {10.1038/s41586-023-06536-0},
	Id = {Park2023},
	Isbn = {1476-4687},
	Journal = {Nature},
	Number = {7981},
	Pages = {74--79},
	Title = {Observation of fractionally quantized anomalous Hall effect},
	Ty = {JOUR},
	Url = {https://doi.org/10.1038/s41586-023-06536-0},
	Volume = {622},
	Year = {2023}}

@article{potter,
  title = {Quantum Spin Liquids and the Metal-Insulator Transition in Doped Semiconductors},
  author = {Potter, Andrew C. and Barkeshli, Maissam and McGreevy, John and Senthil, T.},
  journal = {Phys. Rev. Lett.},
  volume = {109},
  issue = {7},
  pages = {077205},
  numpages = {5},
  year = {2012},
  month = {Aug},
  publisher = {American Physical Society},
  doi = {10.1103/PhysRevLett.109.077205},
  url = {https://link.aps.org/doi/10.1103/PhysRevLett.109.077205}
}

@article{song_2024density,
  title = {Density wave halo around anyons in fractional quantum anomalous Hall states},
  author = {Song, Xue-Yang and Senthil, T.},
  journal = {Phys. Rev. B},
  volume = {110},
  issue = {8},
  pages = {085120},
  numpages = {7},
  year = {2024},
  month = {Aug},
  publisher = {American Physical Society},
  doi = {10.1103/PhysRevB.110.085120},
  url = {https://link.aps.org/doi/10.1103/PhysRevB.110.085120}
}

@article{xia2026,
	Author = {Xia, Yiyu and Han, Zhongdong and Zhu, Jiacheng and Zhang, Yichi and Kn{\"u}ppel, Patrick and Watanabe, Kenji and Taniguchi, Takashi and Mak, Kin Fai and Shan, Jie},
	Da = {2026/01/28},
	Doi = {10.1038/s41586-025-10049-3},
	Id = {Xia2026},
	Isbn = {1476-4687},
	Journal = {Nature},
	Title = {Bandwidth-tuned Mott transition and superconductivity in moir{\'e}WSe2},
	Ty = {JOUR},
	Url = {https://doi.org/10.1038/s41586-025-10049-3},
	Year = {2026}}

@article{crepel2023fci,
  title = {Anomalous Hall metal and fractional Chern insulator in twisted transition metal dichalcogenides},
  author = {Cr\'epel, Valentin and Fu, Liang},
  journal = {Phys. Rev. B},
  volume = {107},
  issue = {20},
  pages = {L201109},
  numpages = {5},
  year = {2023},
  month = {May},
  publisher = {American Physical Society},
  doi = {10.1103/PhysRevB.107.L201109},
  url = {https://link.aps.org/doi/10.1103/PhysRevB.107.L201109}
}

@article{wu2019topological,
  title={Topological insulators in twisted transition metal dichalcogenide homobilayers},
  author={Wu, Fengcheng and Lovorn, Timothy and Tutuc, Emanuel and Martin, Ivar and MacDonald, AH},
  journal={Physical review letters},
  volume={122},
  number={8},
  pages={086402},
  year={2019},
  publisher={APS}
}

@article{devakul2021magic,
  title={Magic in twisted transition metal dichalcogenide bilayers},
  author={Devakul, Trithep and Cr{\'e}pel, Valentin and Zhang, Yang and Fu, Liang},
  journal={Nature communications},
  volume={12},
  number={1},
  pages={6730},
  year={2021},
  publisher={Nature Publishing Group UK London}
}

@article{wilhelm2021interplay,
  title={Interplay of fractional Chern insulator and charge density wave phases in twisted bilayer graphene},
  author={Wilhelm, Patrick and Lang, Thomas C and L{\"a}uchli, Andreas M},
  journal={Physical Review B},
  volume={103},
  number={12},
  pages={125406},
  year={2021},
  publisher={APS}
}

@article{yu2020giant,
  title={Giant magnetic field from moir{\'e} induced Berry phase in homobilayer semiconductors},
  author={Yu, Hongyi and Chen, Mingxing and Yao, Wang},
  journal={National Science Review},
  volume={7},
  number={1},
  pages={12--20},
  year={2020},
  publisher={Oxford University Press}
}

@article{li2021spontaneous,
  title={Spontaneous fractional Chern insulators in transition metal dichalcogenide moir{\'e} superlattices},
  author={Li, Heqiu and Kumar, Umesh and Sun, Kai and Lin, Shi-Zeng},
  journal={Physical Review Research},
  volume={3},
  number={3},
  pages={L032070},
  year={2021},
  publisher={APS}
}

@article{ledwith2020fractional,
  title = {Fractional Chern insulator states in twisted bilayer graphene: An analytical approach},
  author = {Ledwith, Patrick J. and Tarnopolsky, Grigory and Khalaf, Eslam and Vishwanath, Ashvin},
  journal = {Phys. Rev. Res.},
  volume = {2},
  issue = {2},
  pages = {023237},
  numpages = {12},
  year = {2020},
  month = {May},
  publisher = {American Physical Society},
  doi = {10.1103/PhysRevResearch.2.023237},
  url = {https://link.aps.org/doi/10.1103/PhysRevResearch.2.023237}
}

@article{cao2018unconventional,
  title={Unconventional superconductivity in magic-angle graphene superlattices},
  author={Cao, Yuan and Fatemi, Valla and Fang, Shiang and Watanabe, Kenji and Taniguchi, Takashi and Kaxiras, Efthimios and Jarillo-Herrero, Pablo},
  journal={Nature},
  volume={556},
  number={7699},
  pages={43--50},
  year={2018},
  publisher={Nature Publishing Group UK London}
}

@article{cao2018correlated,
  title={Correlated insulator behaviour at half-filling in magic-angle graphene superlattices},
  author={Cao, Yuan and Fatemi, Valla and Demir, Ahmet and Fang, Shiang and Tomarken, Spencer L and Luo, Jason Y and Sanchez-Yamagishi, Javier D and Watanabe, Kenji and Taniguchi, Takashi and Kaxiras, Efthimios and others},
  journal={Nature},
  volume={556},
  number={7699},
  pages={80--84},
  year={2018},
  publisher={Nature Publishing Group UK London}
}

@article{PhysRevB.46.5621,
  title = {Gauge theory of the normal state of high-${\mathit{T}}_{\mathit{c}}$ superconductors},
  author = {Lee, Patrick A. and Nagaosa, Naoto},
  journal = {Phys. Rev. B},
  volume = {46},
  issue = {9},
  pages = {5621--5639},
  numpages = {0},
  year = {1992},
  month = {Sep},
  publisher = {American Physical Society},
  doi = {10.1103/PhysRevB.46.5621},
  url = {https://link.aps.org/doi/10.1103/PhysRevB.46.5621}
}

@article{PhysRevB.52.17275,
  title = {Quantum Boltzmann equation of composite fermions interacting with a gauge field},
  author = {Kim, Yong Baek and Lee, Patrick A. and Wen, Xiao-Gang},
  journal = {Phys. Rev. B},
  volume = {52},
  issue = {24},
  pages = {17275--17292},
  numpages = {0},
  year = {1995},
  month = {Dec},
  publisher = {American Physical Society},
  doi = {10.1103/PhysRevB.52.17275},
  url = {https://link.aps.org/doi/10.1103/PhysRevB.52.17275}
}

@article{PhysRevB.76.235124,
  title = {Transport properties of a spinon Fermi surface coupled to a U(1) gauge field},
  author = {Nave, Cody P. and Lee, Patrick A.},
  journal = {Phys. Rev. B},
  volume = {76},
  issue = {23},
  pages = {235124},
  numpages = {11},
  year = {2007},
  month = {Dec},
  publisher = {American Physical Society},
  doi = {10.1103/PhysRevB.76.235124},
  url = {https://link.aps.org/doi/10.1103/PhysRevB.76.235124}
}

@article{PhysRevB.82.045121,
  title = {Controlled expansion for certain non-Fermi-liquid metals},
  author = {Mross, David F. and McGreevy, John and Liu, Hong and Senthil, T.},
  journal = {Phys. Rev. B},
  volume = {82},
  issue = {4},
  pages = {045121},
  numpages = {19},
  year = {2010},
  month = {Jul},
  publisher = {American Physical Society},
  doi = {10.1103/PhysRevB.82.045121},
  url = {https://link.aps.org/doi/10.1103/PhysRevB.82.045121}
}

\end{document}